\documentclass[%
 reprint,
superscriptaddress,
 amsmath,amssymb,
 aps,
pra,
]{revtex4-2}

\usepackage{graphicx}
\usepackage{dcolumn}
\usepackage{bm}
\usepackage[capitalize]{cleveref}
\usepackage[normalem]{ulem}
\usepackage{xcolor}
\usepackage[makeroom]{cancel}
\usepackage{wasysym}
\usepackage{physics}
\newcommand{\vect}[1]{\boldsymbol{\mathbf{#1}}}

\newcommand{\cjs}[1]{\textcolor{cyan}{\ifmmode\text{\cancel{\ensuremath{#1}}}\else\sout{#1}\fi}}

\newcommand{\bfs}[1]{\textcolor{blue}{\ifmmode\text{\cancel{\ensuremath{#1}}}\else\sout{#1}\fi}}

\newcommand{\jls}[1]{\textcolor{magenta}{\ifmmode\text{\cancel{\ensuremath{#1}}}\else\sout{#1}\fi}}

\newcommand{\TU}{Institute of Physics, Faculty of Physics, Astronomy and Informatics, Nicolaus Copernicus University
in Toru\'n, Grudzi\c{a}dzka 5, 87-100 Toru\'n, Poland }

\begin{document}

\preprint{APS/123-QED}

\title{Spectral functions of the strongly interacting 3D Fermi gas}

\author{Christian H. Johansen}
 \affiliation{Max-Planck-Institut f\"ur Physik komplexer Systeme, 01187 Dresden, Germany}
 \affiliation{\TU}
 \author{Bernhard Frank}
 \affiliation{Institut f\"ur Theoretische Physik and W\"urzburg-Dresden Cluster of Excellence ct.qmat, Technische Universit\"at Dresden, 01062 Dresden, Germany}
\author{Johannes Lang}%
\email[]{j.lang@uni-koeln.de}
\affiliation{Institut f\"ur Theoretische Physik, Universit\"at zu K\"oln, Z\"ulpicher Stra{\ss}e 77, 50937 Cologne, Germany}%
\affiliation{Max-Planck-Institut f\"ur Physik komplexer Systeme, 01187 Dresden, Germany}%

\date{23 January 2024}

\begin{abstract}
Computing dynamical properties of strongly interacting quantum many-body systems poses a major challenge to theoretical approaches. Usually, one has to resort to numerical analytic continuation of results on imaginary frequencies, which is a mathematically ill-defined procedure. Here, we present an efficient method to compute the spectral functions of the two-component Fermi gas near the strongly interacting unitary limit directly in real frequencies. To this end, we combine the Keldysh path integral that is defined in real time with the self-consistent T-matrix approximation. The latter is known to predict thermodynamic and transport properties in good agreement with experimental observations in ultracold atoms. We validate our method by comparison with thermodynamic quantities obtained from imaginary time calculations and by transforming our real-time propagators to imaginary time. By comparison with state-of-the-art numerical analytic continuation of the imaginary time results, we show that our real-time results give qualitative improvements for dynamical quantities. Moreover, we show that no significant pseudogap regime exists in the self-consistent T-matrix approximation above the critical temperature $T_c$, an issue that has been under significant debate. We close by pointing out the versatile nature of our method as it can be extended to other systems, like the spin- or mass-imbalanced Fermi gas, other Bose-Fermi models, 2D systems as well as systems out of equilibrium.
\end{abstract}

\maketitle

\section{Introduction}
The two-component Fermi gas, realized in experiments with ultra-cold atoms both in 2D and 3D, nowadays plays a central role in the study of strongly correlated systems \cite{Bloch2008,Zwerger2012, Zwerger2014}. 
This is a consequence of the simple form of the interactions that only depend on the scattering length. 
However, at the same time the system exhibits a highly non-perturbative quantum many-body regime close to the quantum mechanical unitary scattering limit reached in the vicinity of a Feshbach resonance. 
The presence of only a single relevant interaction parameter gives rise to universal physics (with corrections in 2D \cite{Hofmann2010,Gao2012}). 
Despite this simplicity the low-temperature phase diagram contains a variety of neutral superfluid phases depending on the spin polarization~\cite{Son2006, Giorgini2008}.
Equally important for the scientific interest in the Fermi gas is the fact that it can be investigated experimentally with a high degree of control. 
One of the first examples was the experimental determination of the phase diagram as a function of the spin-polarization \cite{Shin2008}. 
Later, the thermodynamic properties of the spin-balanced gas at the Feshbach resonance were measured with high precision in both the superfluid and normal phase \cite{Ku2012}.
More recently, further technological advances have allowed the direct measurement of the Tan contact \cite{Tan2008a,Tan2008b,Tan2008c} for the balanced Fermi gas at unitarity with a homogeneous box trap \cite{Mukherjee2019}. 
The latter also paved the way for the determination of local transport coefficients like sound diffusion \cite{Patel2020}.

The remarkably high precision reached in experiments in combination with the universal and non-perturbative nature of the system makes it ideal for benchmarking theoretical approaches. 
The self-consistent T-matrix approximation is a theoretical method that has proven surprisingly accurate throughout the BCS-BEC crossover. 
In particular, its predictions for thermodynamic quantities including critical temperature, Bertsch parameter, and the Tan contact agree extremely well with experimental observations \cite{Haussmann1994, Haussmann2007, Enss2011}. 
Furthermore, combining the self-consistent T-matrix approximation with suitable scattering time approximations even describes the transport properties on a quantitative level \cite{Frank2020}.

Experiments not only focus on thermodynamic and transport quantities but also aim for the investigation of dynamical properties.
The latter is achieved using radio-frequency (rf) spectroscopy which provides access to the spectral functions of the system \cite{Stewart2008}.
With this technique, experiments have been able to measure the superfluid gap of an imbalanced gas  \cite{Schirotzek2008}.
The spectral measurements have also revealed the existence of an intermediary regime in a spin-imbalanced Fermi gas where the quasiparticle picture breaks down, manifesting through the increasing width of the spectral features \cite{Yan2019}. 
In a 2D gas, another interesting spectral feature has been measured, namely the existence of a pseudogap above the critical temperature \cite{Feld2011}.
While the existence of a pseudogap was expected in 2D, its fate in 3D is less obvious. Although a typical back-bending of the dispersion has been observed well above the critical temperature \cite{Gaebler2010, Li2023} the expected suppression of the spin susceptibility is absent \cite{Sommer2011}.

Having direct experimental access to the spectral information motivates theoretical computation of the spectrum with equally high reliability as thermodynamic quantities.
It is expected, that self-consistent approximations produce reasonable spectra across all scales without obvious artifacts. 
Combined with the good agreement between experimental thermodynamic observables, we thus expect similar quantitative agreement for the spectral properties obtained through the self-consistent T-matrix approximation.
However, reliably extracting spectral properties with high accuracy has proven difficult, as the procedure to date still involves the numerical analytic continuation (NAC) of data in imaginary frequencies to the real frequency axis, a mathematically ill-defined process, that results in uncontrolled errors \cite{Jarrell1996,Goulko2017}. 
As a result, the comparison of dynamical information gained from rf-spectra lacks the triumphs of determining the thermodynamic and universal parameters \cite{Haussmann2009, Magierski2009, Schmidt2011, Pini2019}.

In this work, we develop a method based on the Keldysh path integral that allows us to calculate the self-consistent T-matrix approximation to very high precision directly in real frequencies. 
We exploit the convolution structure of the T-matrix approximation to express it through Fourier transforms. 
The key to the success of our approach lies in a piecewise interpolation scheme that allows one to analytically treat the fast oscillations that inevitably appear in such a real-time formulation.
The method is validated against previous results for thermodynamic quantities obtained from Matsubara propagators in imaginary time and differs by less than $1\%$.
This level of precision is remarkable because the previous results are known to be very accurate in imaginary time, demonstrating that our method is capable of reliably computing thermodynamic properties. 
For systems in thermal equilibrium, the main advantage of our technique is the direct access to dynamical quantities as opposed to the imaginary-time approach, which yields uncontrolled errors when analytically continued to real frequencies.
As a result, the formulation of the self-consistent T-matrix within Keldysh field theory allows us to quantitatively describe both spectral features and thermodynamics.

The paper is structured as follows, in Sec.~\ref{sec:Model}, we introduce the two-component Fermi gas as realized in experiments with ultra-cold atoms. 
Readers familiar with this subject can skip ahead to Eq.~\eqref{eq:Ham}. 
The Keldysh formalism is recapitulated and applied to the Fermi gas in Sec.~\ref{sec:Formalism}. 
Our central technological development is presented in Sec.~\ref{sec:method} and carefully validated in Secs.~\ref{sec:validation} and \ref{sec:comp}. 
We then apply the method to spectral functions near the superfluid transition in the BCS-BEC crossover, addressing the issue of the pseudogap at unitarity (Sec.~\ref{sec:critical}) before we provide a short outlook in Sec.~\ref{sec:Conclusion}. 

\section{The interacting Fermi gas}\label{sec:Model}
A gas with density $n$, by definition, describes a dilute system, when the average distance between particles $n^{-1/3}$ is much larger than the size of the particles. 
Under these conditions, neutral particles quite generally interact via induced dipole-dipole interactions resulting in an attractive two-body potential generated by the van der Waals force \cite{Chin2010}. 
To describe the two-body scattering potential one can use a simple model, which cuts off the attraction at some small atomic distance $R_0$ \cite{Zwerger2014}
\begin{equation}\label{eq:VdW_pot}
V(r)=\begin{cases}
-C_6/r^6,& r>R_0,\\
\infty, &r\leq R_0,
\end{cases}
\end{equation}
where $C_6$ is a positive constant that is determined by the specific atoms used in the experiments.  

At low temperatures, the gas becomes degenerate, meaning that the particles' wave functions overlap.
This behavior manifests when  $n \lambda_T^3\gg 1$ with $\lambda_T=\hbar\sqrt{2\pi/mT}$ being the thermal de Broglie wavelength, implying that scattering needs to be treated quantum mechanically \cite{Zwerger2014}.  

To simplify the theoretical model one can exploit a separation of length scales in the ultra-cold dilute quantum gas
\begin{equation}\label{eq:lengthScales}
R_0\ll l_{vdW}\ll n^{-1/3}\ll \lambda_T,
\end{equation}
where $l_{vdW}=(mC_6/\hbar^2)^{1/4}/2$ is the length scale associated with the van der Waals interaction in \cref{eq:VdW_pot}. 
Due to the small kinetic energy of the gas, it is sufficient to consider the s-wave scattering channel with the isotropic scattering amplitude \cite{Gottfried2003}
\begin{equation}\label{eq:scatAmp}
f(k)=\frac{-1}{a^{-1}-r_e k^2/2+... +ik},
\end{equation}  
with $a$ being the scattering length and $r_e$ the effective range. 
For the potential in \cref{eq:VdW_pot} the scattering length and effective range can  be analytically computed
 \cite{Zwerger2014,Flambaum1993,Harabati1999}
 \begin{subequations}
 \begin{align}
     a&=\tilde{a}\bigg(1-\tan\left(\Phi-3\pi/8\right)\bigg),\label{eq:a_vdW}\\
     r_e&=2.92 \tilde{a}\left(1-2\frac{\tilde{a}}{a}+2\left(\frac{\tilde{a}}{a}\right)^2\right).\label{eq:vdW}
     \end{align}
 \end{subequations}
The results can be parametrized through the mean scattering length  $\tilde{a}=0.956 l_{vdW}$ and the WKB-phase $\Phi=2l_{vdW}^2/R_0^2$.  
Looking at \cref{eq:a_vdW} it is seen that the scattering length diverges for specific values of $\Phi$. 
Such a divergence appears when the collision of the incoming particles is resonant with a bound state. Even though these resonances are determined by the atomic scales $l_{vdW}$ and $R_0$, Feshbach resonances \cite{Feshbach1958,Fano1961} allow one to tune the scattering properties via an external magnetic field. 

Here we only briefly describe the main concept in the experimentally relevant case of alkali atoms with one valence electron and refer the interested reader to the detailed review \cite{Chin2010}. 
A Feshbach resonance emerges because the external magnetic field polarizes the valence electrons of the atoms. 
For low energies, the physics of a collision is therefore mainly described by the triplet state.  
However, the wave function for the singlet state of two fermions is symmetric, which means that the singlet potential at small distances is much more attractive, compared to the triplet potential, and therefore supports many bound states. 
Because the singlet and triplet states have different magnetic moments, it is possible to create a resonance between the incoming triplet states (open channel) and a bound singlet state (closed channel), by tuning the magnetic field. 
As the magnetic field is finite there will be interactions between the close-to-resonance open channel and the closed channel.
To describe two fermionic atoms of the same type scattering into the open channel, one can use the two-channel model \cite{Timmermans2001,Holland2001}
\begin{align}\label{eq:2Ch_Ham}
\begin{split}
H_{2c}=&\int\dd^3 r\Bigg[\sum_\sigma \psi_\sigma^\dagger(\textbf{r})\left(-\frac{\hbar^2\nabla^2}{2m}\right)\psi_\sigma(\textbf{r})\\&+\Delta^\dagger (\textbf{r})\left(-\frac{\hbar^2\nabla^2}{4m}+\nu(B)\right)\Delta(\textbf{r})\\
&+\!\tilde{g}\!\int\!\!\dd^3 r' \chi(\abs{\textbf{r}-\textbf{r}'})\\ &\qquad\quad\times\!\left(\!\Delta^\dagger\!\left(\frac{\textbf{r}+\textbf{r}'}{2}\right)\psi_\uparrow(\textbf{r})\psi_\downarrow(\textbf{r}')+\text{h.c.}\!\right)\!\Bigg],
\end{split}
\end{align}
where $\psi_\sigma$ and $\psi_\sigma^\dagger$ are the fermionic annihilation and creation operators with spin $\sigma$ and mass $m$. 
The closed channel is described by the annihilation operator of the bosonic pairing field $\Delta$, as well as the corresponding creation operator $\Delta^\dagger$.
The magnetic field $B$ leads to a detuning of the closed channel by $\nu(B)=\delta \mu\left(B-B_c\right)$, with the difference in the magnetic moment between the triplet and singlet states being $\delta\mu$ and $B_c$ denoting the field strength at which they become resonant. 
The scattering of two fermions in the open channel into the pairing field happens through an isotropic form factor that is normalized $\int \dd^3 r\, \chi(\abs{\textbf{r}})=1$. 
This means that the strength of the interaction is given by $\tilde{g}$.
By eliminating the pairing field one can compute the corresponding scattering length and effective range in \cref{eq:scatAmp} originating from the Feshbach resonance \cite{Schmidt2012}
 \begin{subequations}
 \begin{align}
     a^{-1}&=-\frac{m r^\star}{\hbar^2}\nu(B)+\frac{1}{2 \tilde{a}},\label{eq:a_feshbach}\\
     r_e&=-2r^\star+3\tilde{a}\left(1-\frac{4 \tilde{a}}{3 a}\right),\label{eq:r_feshbach}
     \end{align}
 \end{subequations}
where $r^\star=2\pi\hbar ^4 /(m\tilde{g})^2$ is the length scale associated to the Feshbach coupling. 
Importantly there exists a critical field strength $B_0=\hbar^2(2\tilde{a}\,\delta\mu\, m\, r^\star)^{-1}+B_c$ at which the scattering length $a$ diverges.

The behavior of the system qualitatively differs depending on the relation between the van der Waals length scale and the strength of the Feshbach resonance. 
This motivates introducing the resonance strength $s_{res}=\tilde{a}/r^\star$ \cite{Chin2010}, which compares the two scales.
In the following, we will consider a so-called open-channel dominated Feshbach resonance where $s_{res}\gg1$, such that $\tilde{g}$ is large. 
Furthermore, such a resonance directly leads to $r_e$ being on the order of $l_{vdW}$. 
Since the Fermi momentum $k_F\approx n^{1/3}$, the separation of length scales in an ultra-cold gas in \cref{eq:lengthScales} means that $r_e k_F\ll 1$.
The system is therefore well described by considering a contact interaction where $r_e\rightarrow 0$.
By rescaling the field operator $\Delta\rightarrow \Delta/\tilde{g}$ the Hamiltonian \cref{eq:2Ch_Ham} can be rewritten as
\begin{equation}\label{eq:Ham}
\begin{aligned}
H=\int\dd^3 r\Bigg[&\sum_\sigma \psi_\sigma^\dagger(\textbf{r})\left(-\frac{\nabla^2}{2m_\sigma}-\mu_\sigma\right)\psi_\sigma(\textbf{r})\\&-\Delta^\dagger (\textbf{r})\frac{1}{g_0}\Delta(\textbf{r})\\&+\Delta^\dagger\left(\textbf{r}\right)\psi_\uparrow(\textbf{r})\psi_\downarrow(\textbf{r})+\text{h.c.}\Bigg],
\end{aligned}
\end{equation}
where we have now switched to units of $\hbar=1$ and $k_B=1$, which will be used throughout the rest of the paper. 
Furthermore, the system has also been generalized by including different chemical potentials of the two different fermion species $\mu_\sigma$ and mass imbalance. 
We have also defined $g_0=-\tilde{g}^2/\nu(B)$ similarly to \cite{Schmidt2013} and neglected the kinetic energy of the bare $\Delta$-field due to $\tilde{g}$ being large. 
This Hamiltonian quantitatively describes experiments with an open-channel dominated Feshbach resonance, which for example is the case for several of the experiments that have investigated the BCS-BEC crossover \cite{Navon2010,Ku2012}.
The zero-range interactions lead to a divergence and the interactions therefore have to be cut off at the length scale $l_{vdW}$. 
This renormalization connects $g_0$ to the experimental scattering length, which is discussed in App.~\ref{app:Gamma}. 

\section{Field theoretical method}\label{sec:Formalism}
With the effective Hamiltonian for the system defined by \cref{eq:Ham}, we now seek to compute observables. 
The system is theoretically challenging because it is strongly interacting, three dimensional and at finite density. 
For computing thermodynamic properties imaginary-time quantum field theory has been very successful.
From this theory dynamical quantities can be computed using numerical analytical continuation (NAC). 
To avoid this inherently uncontrolled method, while still taking advantage of the successes of quantum field theory, we use the real-time Keldysh path integral approach.

In Sec.~\ref{sec:construction} we will give a brief introduction to the generic structure of the Keldysh path integral.
For the reader already familiar with the theory, it will be sufficient to consider the conventions fixed in \cref{eq:full_matrix_invG,eq:RetDyson,eq:deltaGk}, as these will be used in the remaining parts of the paper.
For a more in-depth discussion of the construction of real-time quantum field theories and its different formulations, the interested reader is referred to existing textbooks and reviews \cite{kamenev2011,Berges2004,RammerRev,morawetz2018,altland2010,Jauho2007}. 

Having described the generic structure of the real-time path integral we then, in Sec.~\ref{sec:Tmat}, apply it to the strongly interacting Fermi gas and derive the real-time formulation of the self-consistent T-matrix approximation.

\subsection{Keldysh path integral}\label{sec:construction}
\begin{figure}
    \centering
    \includegraphics[width=\columnwidth]{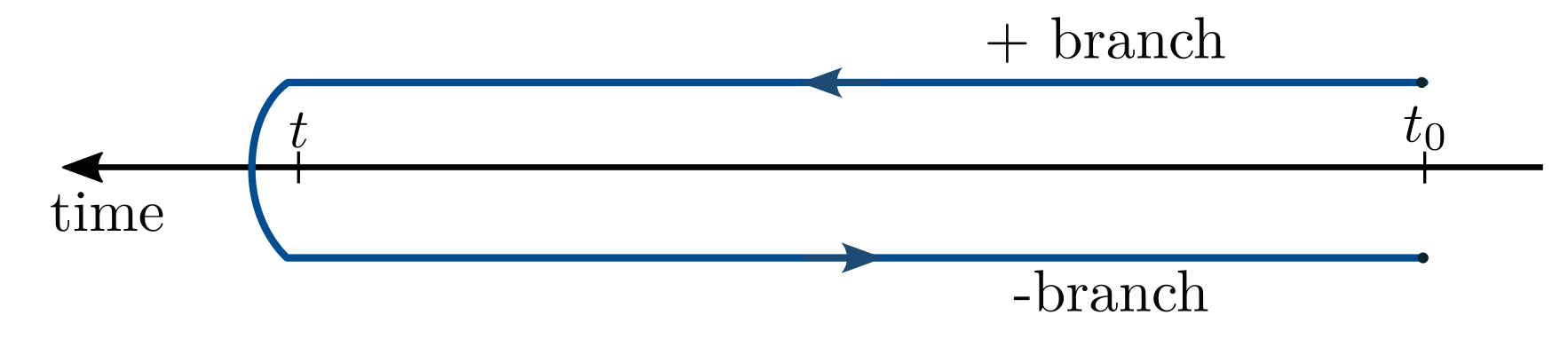}
    \caption{Time contour $\mathfrak{C}$ used to construct a real-time action.}
    \label{fig:contour}
\end{figure}

To construct a quantum field theory in real time, one defines the partition function as a trace over the density matrix
\begin{equation}\label{eq:partFunc}
    \mathcal{Z}=\Tr{\rho(t)}=\Tr{U(t,t_0)\rho(t_0)U(t,t_0)^\dagger},
\end{equation}
where $U(t,t_0)$ is the time-evolution operator. 
With the partition function defined through the density matrix, this formalism allows to consider systems in and out of equilibrium as well as time evolution.
For the current paper we focus on an equilibrium system and use the real-time formulation to avoid the NAC. 
Because $U(t,t_0)$ and $U(t,t_0)^\dagger$ contain opposite time-ordering, it is necessary to differentiate between forward and backward propagation. 
This can conveniently be done using the time contour $\mathfrak{C}$ shown in \cref{fig:contour}.

Using this contour, the system is evolved from some initial time $t_0$ to some time $t$ on the $+$ branch and then back to $t_0$ along the $-$ branch \cite{kamenev2011}.
To evaluate the partition function using the contour $\mathfrak{C}$, one uses coherent Grassmann fields to describe fermionic degrees of freedom and coherent complex fields to describe bosonic degrees of freedom, similar to the imaginary time procedure \cite{altland2010}. 
The differentiation between forward and backward propagation is achieved through the use of two different fields $\phi_{+/-}$, which respectively have support exclusively on the $+$ or $-$ branch in \cref{fig:contour}.
The partition function can then be written as a functional path integral over the fields $\phi_{\pm}$
\begin{equation}\label{eq:ZfuncInt}
    \mathcal{Z}=\int\mathcal{D}\left\{\phi,\bar{\phi}\right\}\text{e}^{i\mathcal{S}}.
\end{equation}
For complex numbers, the bar denotes complex conjugation while for Grassmann numbers it indicates a different, independent Grassmann number. 
The measure in \cref{eq:ZfuncInt} should be understood as a functional measure over the independent fields, $\phi_{\alpha}(\textbf{r},t)$ and $\bar{\phi}_{\alpha}(\textbf{r},t)$ for Grassmann fields, and $\text{Re}\,\phi_{\alpha}(\textbf{r},t)$ and $\text{Im}\,\phi_{\alpha}(\textbf{r},t)$ for complex fields. In both cases $\alpha\in\{+,-\}$ specifies the branch of the Keldysh contour. 
Following this procedure, the continuous action in time $\mathcal{S}$ appearing in \cref{eq:ZfuncInt} takes the form
\begin{equation}\label{eq:Action}
\mathcal{S}=\int\dd^4 x \sum_{\alpha=+,-} \alpha\left( \bar{\phi}_{\alpha}i\partial_t\phi_\alpha-H\left[\bar{\phi}_\alpha,\phi_\alpha\right]\right),
\end{equation}
where $H\left[\bar{\phi}_\alpha,\phi_\alpha\right]$ is the normal-ordered Hamiltonian density with all field creation (annihilation) operators replaced by $\bar{\phi}_\alpha\,(\phi_\alpha)$. 
As the system is time-independent and only has local interactions, all fields are evaluated at the same space-time point $x=(\textbf{r},t)$. Generalizing the construction to time-dependent Hamiltonians and non-local interactions is straightforward. 
For this action the contour has been extended to the entire real axis and interactions are switched on adiabatically in the infinite past \cite{altland2010}.
For interacting systems it is also convenient to write the action in the form 
\begin{equation}
    \mathcal{S}=\mathcal{S}_0+\mathcal{S}_I,
\end{equation}
where $\mathcal{S}_0$ is quadratic in the fields and $\mathcal{S}_I$ contains the interactions.

The continuous representation \cref{eq:Action} does not explicitly contain the two boundaries at $\pm \infty$ of the contour in \cref{fig:contour}.  
These can be conveniently included after a unitary linear transformation on the contour degrees of freedom \cite{kamenev2011}.
The linear transformation then maps to the symmetric and anti-symmetric superpositions $\phi_{c/q}=\left(\phi_+\pm\phi_-\right)/\sqrt{2}$.
When $\phi$ is a complex number, therefore representing a bosonic degree of freedom, one can derive mean-field equations where $\phi_c$ will represent the possibly finite mean-field. Conversely, in a mean-field description of a bosonic system, $\phi_q$ can be connected to quantum fluctuations. 
To this extent $\phi_c$ is often denoted the classical field and $\phi_q$ is denoted as the quantum field.
For fermionic degrees of freedom, where $\phi$ is a Grassmann number, we only use these conventions to make the equivalence of the microscopic formulation for fermions and bosons apparent.

Considering a spatially homogeneous equilibrium theory, means that the non-interacting action is diagonalized by a transformation to reciprocal space, represented by the four-vector $p=(\textbf{k},\omega)$. 
It is related to real space by the Fourier transformation
\begin{equation}
    f(x)=\int \frac{d^4 p}{(2\pi)^4}\text{e}^{i p\cdot x} f(p),
\end{equation}
where the four-vector dot product is defined as  $p\cdot x=\textbf{k}\cdot\textbf{r}-\omega t$ and $d^4 p=\dd \omega \,\dd \textbf{k}^3$.
In reciprocal space, the bare part of the action is given by
\begin{equation}\label{eq:Spsi}
    \mathcal{S}_0=\int \frac{d^4 p}{(2\pi)^4} \mqty(\bar{\phi}_{c} (p)\\\bar{\phi}_{q}(p))^T\mathcal{G}_{0}^{-1}(p)\mqty(\phi_{c}(p)\\\phi_{q}(p)),
\end{equation}
where the inverse contour propagator takes the form
\begin{equation}
   \mathcal{G}_{0}^{-1}(p)= \mqty(0&(G_{0}^{A})^{-1}(p)\\(G_{0}^{R})^{-1}(p)&P^K(p)).
\end{equation}
The lower off-diagonal element, $(G_{0}^{R})^{-1}(p)$, is the inverse bare retarded propagator which is causal in the sense that
$G_{0}^{R}(x)\sim \theta(t)$.
Consequently, it is connected to the inverse, bare advanced propagator $(G_{0}^{A})^{-1}(p)$ by complex conjugation. 
The lower-diagonal component, $P^K$, is a purely imaginary and infinitesimally small term that ensures the normalization of the partition function.

The effect of interactions, on correlation functions, are included through the self-energy matrix $\underline{\Sigma}$, which has a similar causality structure as the inverse propagator $\mathcal{G}_{0}^{-1}$, such that the interacting inverse contour propagator is
\begin{equation}\label{eq:full_matrix_invG}
    \mathcal{G}^{-1}=\mqty(0&(G_{0}^{A})^{-1}-\Sigma^A\\(G_{0}^{R})^{-1}-\Sigma^R&-\Sigma^K).
\end{equation}
Note, that the bare infinitesimal $P^K$ has been dropped since interactions generally makes $\Sigma^K$ finite. 
Inverting \cref{eq:full_matrix_invG} one arrives at the contour propagator
\begin{equation}\label{eq:full_matrix_G}
    \mathcal{G}=\mqty(G^K&G^{R}\\G^{A}&0).
\end{equation}
This inversion leads to the Dyson equation for the retarded propagator
\begin{equation}\label{eq:RetDyson}
 G^{R}(p)=-i\left\langle \phi_{c}(p)\bar{\phi}_{q}(p)\right\rangle =\frac{1}{\left(G_{0}^{R}\right)^{-1}(p)-\Sigma^R(p)},
\end{equation}
and a similar equation for $G^A$.
Transforming back to real-space and the $\pm$-basis, one recovers the operator definition of the retarded propagator
\begin{equation}\label{eq:RetOperator}
 G^{R}(x,x')=-i\theta(t-t')\left\langle \left[\hat{\phi}(x),\hat{\phi}^\dagger(x')\right]_\mp\right\rangle,
\end{equation}
where $\hat{\phi}$ denotes the field operator and the commutator (minus-sign) is used for bosonic fields while the anti-commutator (plus sign) is used for fermionic fields.
In this basis the relation between the advanced and retarded propagators are $G^A(x,x')=\bar{G}^R(x',x)$.
The retarded propagator contains information about the spectrum of the system, as its imaginary part is directly linked to the spectral function 
\begin{equation}\label{eq:A}
  \mathcal{A}(p)=- 2\Im G^R(p).  
\end{equation}
Probability conservation $\mathcal{Z}\equiv 1$ guarantees that the lower diagonal entry of \cref{eq:full_matrix_G} vanishes. 
Explicitly, one has $\left\langle \phi_{q}(p)\bar{\phi}_{q}(p)\right\rangle=0$.
The upper-diagonal element in $\mathcal{G}$ is referred to as the Keldysh propagator and given by
\begin{equation}
    G^K(p)=-i\left\langle \phi_{c}(p)\bar{\phi}_{c}(p)\right\rangle=\abs{G^R(p)}^2\Sigma^K(p),
\end{equation}
which is anti-Hermitian and related to how the spectrum is occupied \cite{kamenev2011}.
In analogy to \cref{eq:RetOperator} for the retarded propagator, the Keldysh propagator can be connected to the operator definition by transforming to real-space and $\pm$-basis
\begin{equation}\label{eq:KelOperator}
    G^K(x,x')=-i\left\langle \left[\hat{\phi}(x),\hat{\phi}^\dagger(x')\right]_\pm\right\rangle,
\end{equation}
where the upper (lower) sign represents bosonic (fermionic) fields. 
It is often convenient to separate out the vacuum term of $G^K$ and define the occupied part of the spectrum as
\begin{equation}\label{eq:deltaGk}
    \delta G^K(p)=G^K(p)+i\mathcal{A}(p)=\abs{G^R(p)}^2\delta\Sigma^K(p),
\end{equation}
where $\delta \Sigma^K(p)=\Sigma^K(p)-2i \Im\Sigma^R(p)$. 
With this decomposition one can compute the momentum distribution by integrating $\delta G^K$ over frequency
\begin{equation}\label{eq:n_k}
   n(\textbf{k})=\pm i\int \frac{\dd \omega}{4\pi} \delta G^K_\sigma(p),
\end{equation}
where the sign is determined by the statistics of the degree of freedom as in \cref{eq:KelOperator}.
For degrees of freedom with well-defined particle number the density $n$ follows directly by integrating over momentum.

Out of equilibrium, $\delta G^K$ is independent of $G^R$, but in thermal equilibrium, they are linked by the fluctuation-dissipation relation (FDR) 
\begin{equation}\label{eq:FDR}
    i\delta G^K(p)= \pm 2 n_{B/F}(\omega) \mathcal{A}(p),
\end{equation}
with the same sign conventions as in \cref{eq:n_k} and the thermal distribution functions $n_{B/F}(\omega)=\left(\text{e}^{\beta \omega}\mp 1\right)^{-1}$.
Note that in equilibrium $\delta\Sigma^K$ and $\Im \Sigma^R$ also satisfy the FDR~\eqref{eq:FDR}.
This means that in thermal equilibrium, similar to the formulation in imaginary time, all interaction effects are encoded in a single function: the retarded self-energy.
However, as discussed in the appendix, computing the Keldysh self-energy directly can be advantageous numerically as it is generally less sensitive to numerical cutoffs in frequency and momentum space than $G^R$. The FDR can therefore be used to remove numerical noise from the spectral function. 

\subsection{Self-consistent T-matrix}\label{sec:Tmat}
Having outlined the generic construction of the Keldysh field theory we now apply it to the strongly interacting Fermi gas.
The Hamiltonian in \cref{eq:Ham} contains two different fermionic degrees of freedom, which we represent with the Grassmann fields $\psi_{c/q,\sigma}$ with $\sigma\in\{\uparrow,\downarrow\}$. 
The bosonic pairing field on the other hand is represented with the complex fields $\Delta_{c/q}$. 
Following the procedure in the previous section the action can be split into three parts
\begin{equation}\label{eq:Sdecomp}
    \mathcal{S}=\mathcal{S}_{\psi}+\mathcal{S}_{\Delta}+\mathcal{S}_{I}.
\end{equation}
The action of free fermions $\mathcal{S}_\psi$ is described by the inverse retarded/advanced propagators
\begin{equation}
    (G_{0,\sigma}^{R/A})^{-1}(p)=\omega-\vect{k}^2/2m_\sigma+\mu_\sigma\pm i 0^+,
\end{equation}
while the inverse free propagators for the pairing field are
\begin{equation}
    \left(\Gamma_0^{R/A}\right)^{-1}=1/g_0\pm i 0^+.
\end{equation}
In both cases the infinitesimal imaginary part ensures the correct causal form.
When deriving the free action for the pairing field $\mathcal{S}_\Delta$, the dynamic part has been neglected.
This is consistent with the contact interaction approximation which requires a renormalization of the coupling $g_0$. 
In that procedure $g_0\rightarrow0$ simultaneously with a UV cutoff that goes to $\infty$ in such a way, that the experimental binding energy is recovered.
This procedure means $\omega\ll 1/g_0$ for all physically relevant values of $\omega$, and the $\omega$-dependence can therefore be neglected.

The final term of the action in \cref{eq:Sdecomp} describes interactions between the fermions and the pairing field
\begin{equation}\label{eq:Sint}
\begin{aligned}
    S_I=&-\frac{1}{\sqrt{2}}\int \!\prod_{i=1}^3 \frac{d^4 p_i}{(2\pi)^4} \Big[M^{\gamma}_{\alpha,\beta} \bar{\psi}_{\alpha,\uparrow}(p_1)\bar{\psi}_{\beta,\downarrow}(p_2)\Delta_{\gamma}(p_3)\\&\qquad\qquad\qquad\qquad\times\delta(p_3-p_1-p_2)+c.c\Big],
\end{aligned}
\end{equation}
where the sum over Keldysh indices $\{\alpha,\beta,\gamma\}\in\{c,q\}$ is implicit but the four-momentum structure has been kept explicit. 
The components of the tensor $M$ are given by the first Pauli matrix $M^{c}_{\alpha,\beta}=\sigma^1_{\alpha,\beta}$ and $M^q_{\alpha,\beta}=\delta_{\alpha,\beta}$. 

The inclusion of interactions necessitates finding a suitable approximation scheme for the self-energies.
To compute thermodynamic quantities that satisfy all thermodynamic relations, it is necessary to employ a self-consistent treatment of the pairing channel \cite{Haussmann2007, Frank2018, Pini2019}.
To derive such an approximation one can start from the the 2-particle-irreducible effective action \cite{Baym1962, Cornwall1974} or a similar formulation using the Luttinger-Ward functional $\Phi$ \cite{Luttinger1960}, which is the interaction part of the grand canonical potential. 
The formulation in terms of the former is more common in the context of Keldysh field theory, as it is more general, applying also when the system is out of equilibrium. 
There the self-consistent approximation preserves all conservation laws of the exact action, wherefore it is called a conserving approximation in this context. 
Explicitly for the Hamiltonian in \cref{eq:Ham}, this entails the conservation of particle number, energy, and momentum.
We point out that in the present equilibrium setting both formalisms are identical. 
To relate to the existing literature we express our approximation in terms of the $\Phi$-functional augmented by the Keldysh structure of the propagators.

\begin{figure}
    \centering
    \includegraphics[width=0.9\columnwidth]{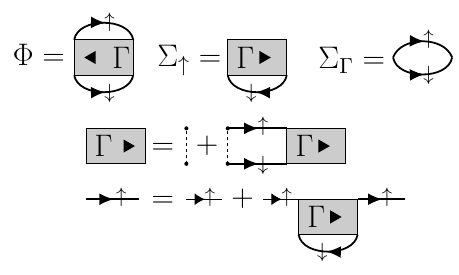}
    \caption{The $\Phi$-functional, the resulting self-energies, and the Dyson equations for both the pairing field and spin-$\uparrow$ fermion. The full fermion propagators are black bold lines, with an arrow indicating the species and the pair propagator is the grey box labeled with $\Gamma$. The arrows indicate the direction of propagation and dashed line is the bare contact interaction for the electrons, while the thin fermion lines represent the bare fermion propagators. The self-energy and Dyson equation for $\downarrow$ is obtained by flipping the spins of the spin-$\uparrow$ counterparts.}
    \label{fig:Phi}
\end{figure}

The $\Phi$-functional is elegantly represented diagrammatically as a sum of closed diagrams constructed from the vertices of the action, such that in each diagram more than two lines have to be cut for it to become disconnected. 
For the interaction in \cref{eq:Sint} the 
lowest order $\Phi$-functional that can be constructed is the two-loop diagram shown in \cref{fig:Phi}. 
This forms the basis of the self-consistent T-matrix approximation.
Due to the contact nature of the interaction, the Bethe-Salpeter equation in \cref{fig:Phi} has the same form as the Dyson equation, which is the reason why the pairing field description is convenient. 
The self-energies can be derived by taking functional derivatives of $\Phi$ with respect to the propagators
\begin{equation}\label{eq:LW}
    \Sigma_\sigma=\frac{\delta\Phi}{\delta G_\sigma} \text{ and } \Sigma_\Gamma=\frac{\delta\Phi}{\delta \Gamma}.
\end{equation}
The resulting self-energies are also drawn in \cref{fig:Phi}, where all propagators are understood as dressed by self-energies. 
Hence, \cref{eq:LW} leads to three coupled self-consistent integral equations for the retarded self-energies, which together with the three corresponding Dyson equations of the form of \cref{eq:RetDyson} give rise to a closed set of equations. 
At this point, the general structure of the theory has been fully defined, recovering the self-consistent T-matrix approximation \cite{Haussmann1994} but formulated directly in real frequencies.

By comparison with high-precision experiments \cite{Ku2012,Mukherjee2019} and bold diagrammatic Monte-Carlo results \cite{VanHoucke2012,Rossi2018b} this approximation has proven surprisingly accurate for thermodynamic quantities in the experimentally relevant regime of strong interactions $|k_F a|\gtrsim 1$. 
Outside this regime larger deviations from exactly known results occur: for weak interactions, BCS theory is recovered, lacking the quantitatively important corrections found by Gor`kov and Melik-Barkhudarov \cite{Gorkov1961}. 
For very deeply bound pairs, the gas is well described in terms of weakly interacting bosons. 
However, the scattering length between these pairs~\cite{Petrov2004} $a_\text{pp}=0.6 a$ is not correctly recovered by the T-matrix approximation \cite{Haussmann1994}. 
Focusing on parameters near unitarity, several other approximation schemes have been developed. 
These include the historic non-selfconsistent approach by Nozi\`eres and Schmidt-Rink \cite{Nozieres1985}, an expansion in the inverse number of fermionic components \cite{Veillette2007}, the Chevy ansatz \cite{Chevy2006}, Gaussian fluctuations \cite{Melo1993,Diener2008} as well as partially self-consistent T-matrix approximations \cite{Pini2019}. 
Comparison across these methods systematically shows that thermodynamic quantities are most accurately approximated by the self-consistent T-matrix.

In contrast to an imaginary-time formulation, self-energies are not only distinguished by particle species but also by their Keldysh index $R/A/K$. 
Hence, one has to account for the contour degree of freedom ($q, c$), which we have suppressed in our discussion of \cref{eq:LW}. 
Its inclusion is straightforward: One constructs all diagrams that involve an odd number of quantum fields at each vertex, thereby accounting for the fact that $S_\text{int}$ is antisymmetric under exchange of the $+$ and $-$ branches of the Keldysh contour $\mathfrak{C}$.
The contraction of two legs is then identified as one of the four possible propagators $\left\langle \phi_{\alpha}(p)\bar{\phi}_{\beta}(p)\right\rangle$, with $\alpha,\beta\in\{c,q\}$ and $\phi\in\{\psi_\sigma,\Delta\}$ thus removing all diagrams that contain a $q$-$q$ propagator.  

After extracting the vacuum contribution as discussed previously, one finds a retarded self-energy for the pair propagator given by
\begin{equation}\label{eq:PP_RetSE}
\begin{aligned}
    \Sigma_\Gamma^R(p)=\frac{i}{2}\int \frac{d^4 p'}{(2\pi)^4}\Big(&2G^R_\uparrow\left(p-p'\right)G^R_\downarrow\left(p'\right)\\&+\delta G^K_\uparrow\left(p-p'\right)G^R_\downarrow\left(p'\right)\\&+G^R_\uparrow\left(p-p'\right)\delta G^K_\downarrow\left(p'\right)\Big)\,.
    \end{aligned}
\end{equation}
The first term determines the vacuum limit as it is independent of $\delta G^K$. 
The two remaining terms describe corrections due to occupation in either of the two fermion species. 
In appendix \ref{app:Gamma} we discuss how the vacuum term can be used to renormalize the bare divergent propagator and connect the theory to the experimental scattering length. 
As expected, the Keldysh self-energy describes the occupation of the fluctuations in the pair propagator and thus requires finite occupation of both fermion species 
\begin{equation}\label{eq:PP_SEK}
    \delta \Sigma_\Gamma^K(p)=\frac{i}{2}\int \frac{d^4 p'}{(2\pi)^4} \delta G^K_\uparrow\left(p-p'\right)\delta G^K_\downarrow\left(p'\right).
\end{equation}

The full pair propagator $\Gamma$ is then used in the retarded self-energy for the fermions given by
\begin{equation}\label{eq:fermion_SER}
\begin{aligned}
 \Sigma^R_\sigma\left(p\right)=-\frac{i}{2 }\int \frac{d^4 p'}{(2\pi)^4}\Big(&\Gamma^R\left(p+p'\right) \delta G^K_{\bar{\sigma}}\left(p'\right)\\&+\delta \Gamma^K\left(p+p'\right) G^{A}_{\bar{\sigma}}\left(p'\right)\Big),
\end{aligned}
\end{equation}
where $\bar{\sigma}$ is the opposite species of $\sigma$. 
Since pairs are absent in the vacuum limit, $\Sigma^R_\sigma$ is non-zero only in the presence of a finite density of fermions with opposite spin. This is also the case for the occupation of the fermion spectrum which is determined through \cref{eq:deltaGk} with
\begin{equation}\label{eq:fermion_SEK}
\begin{aligned}
    \delta\Sigma^K_\sigma\left(p\right)=-\frac{i}{2 }\int \frac{d^4 p'}{(2\pi)^4}\delta \Gamma^K\left(p+p'\right)\Big(&\delta G_{\bar{\sigma}}^K\left(p'\right)\\&-2i\mathcal{A}_{\bar{\sigma}}\left(p'\right)\Big).
    \end{aligned}
\end{equation}
The full theory is then defined by the self-energies \cref{eq:PP_RetSE,eq:PP_SEK,eq:fermion_SER,eq:fermion_SEK} dressing the bare propagators of fermions and pairing field and the corresponding Dyson equations in \cref{eq:RetDyson,eq:deltaGk}. 

\section{Numerical implementation}\label{sec:method}
Having formulated the theoretical framework, we now focus on the numerical solution of the coupled self-consistent integral equations for the propagators.

As the retarded and Keldysh propagators are simultaneously diagonal in $p$-space, the inversion of the Dyson equations and the computation of the Keldysh propagator are straightforward.
The main challenge lies in computing the self-energies self-consistently.
It is possible to do this efficiently and accurately thanks to the convolution nature of \cref{eq:PP_RetSE,eq:PP_SEK,eq:fermion_SER,eq:fermion_SEK}.
This means that the self-energies can be obtained by transforming the propagators into the position and time domain then computing their products and transforming these back to momentum and frequency space.

However, performing these transformations accurately is difficult even in imaginary time, because the correlation functions decay slowly at high momenta and frequencies. 
For instance, the momentum distribution, according to the Tan energy theorem~\cite{Tan2008b}, acquires the form $n(\mathbf k) = \mathcal C/\mathbf k^4$. This implies that the momentum-integrated RF spectrum 
$I(\omega) = i \int_\mathbf{k} \delta G^K(\mathbf k,\mathbf{k}^2/2m -\mu -\omega)/2$ scales like $\omega^{-3/2}$ for large frequencies in case of negligible final-state interactions~\cite{Braaten2010} in agreement with the experimental observations~\cite{Stewart2008}. Furthermore, the fermionic spectral functions themselves show power-law asymptotics, which are given below in \cref{eq:high_tail,eq:low_tail}.
Even in imaginary time, it turns out that using a fast Fourier transform and hence an equidistant grid, to discretize propagators, is unfeasible.
Instead it has been shown that using adaptive non-equidistant grids one can solve the imaginary time problem with high accuracy \cite{Haussmann2007, Frank2018}.

In a real-time formulation the need for long grids is much more challenging than in imaginary time.
This is because the high-momentum behaviour of the propagators is quickly oscillating with a frequency that grows linearly with the momentum. 
This so-called chirped oscillation creates the predicament that a direct implementation requires an increasingly dense grid for larger momentum. 
In real-time one therefore has to use long grids which are simultaneously dense, making the direct implementation intractable.

Here we solve this problem by constructing a method which, through a series of different interpolations, analytically separates out the chirped oscillations such that numerical sampling of the latter is completely avoided.
As a result, similar to the calculation on Matsubara frequencies, one can use grids that sparsely sample high momenta such that the power-law tails of the correlations are accurately described. 
The method, therefore, allows one to compute convolutions both efficiently and accurately, such that the self-consistent T-matrix can be computed directly in real time without NAC.

To understand the procedure it is insightful to consider the simplest case where a chirped oscillation appears, namely the transformation of the bare retarded propagator from $p$ to $x$
\begin{equation}
    G^R_0(x)=\int \frac{d^4p}{(2\pi)^4}\frac{\exp(ip\cdot x)}{\omega+\mu-\frac{\textbf{k}^2}{2m}+i0^+}.
\end{equation}
The frequency integral can be computed by contour integration and leads to the propagator in $(\mathbf{k},t)$ space
\begin{equation}\label{eq:G0_kt}
    G^R_0(\vect{k},t)=-i\theta(t) \exp[\left(-i \frac{\textbf{k}^2}{2m}+i\mu - 0^+\right)t],
\end{equation}
which, due to isotropy only depends on $\abs{\textbf{k}}$. 
The first term in the argument of the exponential in \cref{eq:G0_kt} leads to oscillations with a linear chirp in momentum, which is exceedingly hard to sample on a finite numerical grid for larger values of $\abs{\textbf{k}}$.
However, for the bare propagator, the momentum integral can be performed analytically 
\begin{equation}\label{eq:bareG_rt}
    G^R_0(\textbf{r},t)=\theta(t)\exp\left(i\frac{\textbf{r}^2 m}{2t}+(i\mu -0^+)t\right)\!\left(\frac{i m}{2\pi t}\right)^{\!3/2},
\end{equation}
where the oscillations due to $\textbf{k}^2 t$ translates into oscillations with an argument $\textbf{r}^2/t$. 
This is equally hard to sample and both the oscillations and the entire propagator diverge at $t=0$.  
The challenge to be solved thus lies in the efficient and accurate computation of Fourier transforms without relying on sampling these fast oscillations.
To explain our method we will consider the transformation $p\rightarrow x$ and discuss its relation to the inverse transformation at the end of the section.  

\begin{figure}
    \centering
    \includegraphics[width=\columnwidth]{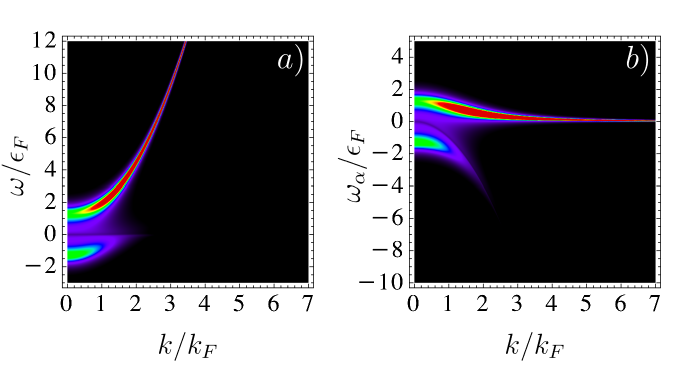}
    \caption{Typical spectral function of an isotropic strongly interacting system with both broadening and a non-trivial structure at small momenta.  
    The bare quadratic dispersion leads to a large change of the energy at large momentum in the original coordinate system shown in a). Performing the coordinate transformation makes the energy constant for large momentum as illustrated in b).}
    \label{fig:coorTrans}
\end{figure}

Our method to perform the transformations relies on the observation that, in the ultraviolet, propagators are only weakly modified by the effects of a finite density. 
Hence, one can expect well-defined quasiparticles with a dispersion mostly determined by the bare mass, which in the case of the pair propagator is $m_\Gamma=m_\uparrow+m_\downarrow$. 
More formally, we require that, by changing frequency coordinates to the bare dispersion, the ultraviolet parts of functions are almost constant. 
To this extent, the frequency coordinate is transformed according to $\omega_\alpha=\omega+\textbf{k}^2 v_\alpha+\mu_\alpha$, where for the transformation to real-space we have $v_\alpha = 1/2m_\alpha$ with $\alpha \in \{\Gamma,\uparrow,\downarrow\}$.
On the transformed coordinates our functions $f_\alpha(\textbf{k},\omega_\alpha)$ are now slowly varying as a function of momentum as illustrated in \cref{fig:coorTrans}. 

Additionally, for long-lived quasiparticles or near the critical point the functions can become very strongly peaked in $\omega_\alpha$ which, due to the frequency transformation,  can be solved by locally and adaptively increasing the density of the $\omega_\alpha$-grid. 
Having denser grids only locally is possible, because $f_\alpha$ varies slowly with $k$, such that the same adaptations to the frequency grid are useful across a large range of momenta.
Performing a numerical Fourier transform on an irregular grid, with $N$ points, scales as $N^2$ compared to $N\log\left(N\right)$ for an equidistant or logarithmic grid \cite{FFT1965,Haines1988,Lang2019}.
To mitigate having to use smaller grids, the temporal Fourier transform is done by using a spline interpolation \cite{Haussmann2007, Frank2018}. 
For a system with rotational invariance, the angular directions of the momentum integral can be solved exactly which leaves a one-dimensional integral for each time (space) grid point $t_j$ ($r_i$), which for a three dimensional system takes the form
\begin{equation}
\begin{aligned}
   f_\alpha(r_i,t_j)=&\text{e}^{i\mu_\alpha t_j}\int_0^{\infty}\frac{\dd k }{(2\pi)^2}\frac{2 k \sin\left(k r_i\right)}{r_i}\\&\times f_\alpha(k,t_j)\exp(-i v_\alpha k^2t_j), 
\end{aligned}
\end{equation}
where from now on a notation convenient for an isotropic system $k=\abs{\textbf{k}}$ and $r=\abs{\textbf{r}}$ will be used.
As the fast oscillation $v_\alpha k^2$ has been extracted by the transformation to $\omega_\alpha$, $f_\alpha$ can be interpolated with a spline of order $M$.
The $\sin\left(k r_i\right)$ factor is not included in the interpolation for two reasons:
The first is that it oscillates quickly for large positions and therefore requires a dense $k$-grid for it to be efficiently interpolated with a spline.
Secondly, it would lead to the spline coefficient being a higher-rank tensor and would greatly increase the memory requirements.
To circumvent these issues we only do a low-order spline (cubic or quintic) on $k f_\alpha$ given by
\begin{equation}
    k f_\alpha(k,t_j)\approx\sum_{l=0}^M a^l_{\alpha;n,j}(k-k_n)^l, \text{ where } k_n\leq k<k_{n+1},
\end{equation}
leading to the real-space representation
\begin{equation}\label{eq:kSplineCoef}
    f_\alpha(r_i,t_j)\approx\sum_{n=0}^{N_k-1}\sum_{l=0}^M a^l_{\alpha;n,j}W_{\alpha;i,j,n}^l,
\end{equation}
where $N_k$ is the number of grid points in the momentum grid and the tensor $W$ is given by
\begin{equation}
W_{\alpha;i,j,n}^l=\int_{k_n}^{k_{n+1}}\frac{\dd k}{\left(2\pi\right)^2}  \frac{2\sin( k r_i)}{r_i} \text{e}^{-i v_\alpha t_j k^2} \left(k-k_n\right)^l.
\end{equation}
To make the calculations tractable this tensor has to be factorized. 
To keep the evaluation of the $k^2$ oscillation exact we factorize $\sin(k r_i)/r_i$ by using a 2-point Hermite interpolation to order $P$.
Such an interpolation guarantees that both the function values and the first $P-1$ derivatives are correct at every point. 
It leads to the approximation \cite{Atkinson1989}
\begin{equation}
2\sin(k r_i)/r_i\approx\sum_{p=0}^{2P-1}b_{i,n}^p \left(k-k_n\right)^{p-c(p)}\left(k-k_{n+1}\right)^{c(p)}, 
\end{equation}
for $k_n\leq k <k_{n+1}$.
Here $c(p)=\lfloor p/(2P)\rfloor \left(p\%(2P-1)\right)$, where $\lfloor\cdot \rfloor$ and $\%$ are the floor and modulo operators respectively.
To ensure that the oscillations for large values of $r$ are correctly captured, while still keeping the density of the $k$-grid manageable, one can choose $P$ based on an upper bound of the error discussed in appendix \ref{app:order}.  
The coefficients $b_{i,n}^p$ can be efficiently computed through the generalized divided differences \cite{gdd} and are independent of the frequency grid.
With this factorization, the tensor $W$ takes the form
\begin{equation}
    W^l_{\alpha;i,j,n}=\sum_{p=0}^{2P-1}b^p_{i,n}I^{l+p}_{\alpha,n,j},
\end{equation}
where the elements in $I^{l+p}_{\alpha,n,j}$ only depend on the index $\alpha$ describing the quickly oscillating factor of the function to be transformed.
Through these interpolations, we are no longer relying on sampling linear chirped oscillations because the elements 
\begin{equation}
\begin{aligned}
    I^{l+p}_{\alpha,n,j}=\sum_{q=0}^{c(p)}z_{n,p,q}\text{e}^{-i v_\alpha t_j k_n^2}\int_{0}^{\Delta k_n}\frac{\dd k}{\left(2\pi\right)^2} &\text{e}^{-i v_\alpha t_j \left(k^2-2 k k_n\right)} \\\times&k^{l+p-c(p)+q},
    \end{aligned}
\end{equation}
can be computed exactly. 
The integrals depend on the time/momentum grids and the bare mass, 
while the coefficient 
\begin{equation}
    z_{n,p,q}=(-1)^{c(p)-q}\mqty(c(p)\\q)\Delta k_n^{c(p)-q},
\end{equation}
only depend on the grid-spacing in momentum $\Delta k_n=k_{n+1}-k_n$.
With these elements the real-space representation is 
\begin{equation}
    f_\alpha(r_i,t_j)\approx\sum_{p=0}^{2P-1}\sum_{l=0}^M \bigg(\mathbf{B}^p\cdot \Big(\mathbf{I}^{p+l}_\alpha\circ\left(\mathbf{A}^l\cdot \mathbf{f}_\alpha \right)\Big)\bigg)_{i,j},
\end{equation}
where $\cdot$ is a matrix contraction and $\circ$ is the element-wise (Hadamard) product. 
The elements of the matrix $\textbf{B}^p$ are $b^p_{i,n}$, while the matrix $\textbf{I}^{p+l}_\alpha$ is build up of $I_{\alpha,n,j}^{p+l}$.
The matrix
$\textbf{A}^l$ is independent of $\alpha$ and leads to the spline coefficients $a^l_{\alpha;n,j}$ in \cref{eq:kSplineCoef} once contracted with the matrix built by $\left(\textbf{f}\right)_{\alpha;m,j}=f_\alpha(k_m,t_j)$.
The first contraction scales as $M\times
N_k^2\times N_t$, whereas the Hadamard step scales as $(2P+M)\times N_k\times N_t$. 
The last contraction scales with $2P\times N_r\times N_k\times N_t$.
As $N_i\gg M,P$ this transformation scales the same way as a naive DFT apart from an overhead of $2P$, or $M$ depending on the relations between the grid lengths. 
We point out that, since $f(k,\omega_\alpha)$ is slowly varying, $M=3$ is sufficient for the present application. 
Furthermore, typical values for $P$, as determined by the condition in App.~\ref{app:order}, fall between $9$ and $18$ for momentum grids with a maximum length of $k\sqrt{\beta}=50$.

An important aspect of the transformation procedure is that $\textbf{B}^n$, $\textbf{I}^n_\alpha$ and $\textbf{A}^n$ only depend on the $r,k$ and $t$ grids and $v_\alpha$, which means that all tensors can be precomputed at initialization and are unaffected by the adaptive frequency grid.
To reliably transform all functions to $r$ we need three different coordinate transformations with the choices $v_\alpha\in \{1/m_\Gamma,\,1/m_\uparrow,\,1/m_\downarrow\}/2$.
For transforming from $r$ to $k$, the transformation is identical apart from the substitutions $r\leftrightarrow k$, $v_\alpha t \rightarrow -1/(v_\alpha t)$ and the factor of $\left(2\pi\right)^2$ in $\textbf{I}_\alpha^n$.

\subsection{Comparison to conventional FFT}
The alternative implementation using a naive fast Fourier transform (FFT) would require one to be able to sample the fast oscillations from the bare dispersions. 
For the transformation to space, one has to sample the oscillating factor $\text{e}^{-i k^2 t}$, where for simplicity we have chosen units with $m=1/2$.  By expanding the argument around the largest $k$ and $t$ value it becomes clear that one needs to be able to sample an oscillation with a frequency of  $2 k_{max} t_{max}$.  
To satisfy the Nyquist criterion the momentum grid has to have a spacing smaller than $\Delta_k=\pi /(2k_{max} t_{max})$.
Using a conventional FFT such a grid spacing requires that $r_{max}=2\pi/\Delta_{k}$.
However, the same applies to the inverse transformation from position to momentum space. 
There one has to be able to sample oscillations of the form $\text{e}^{i r^2/t}$. 
Again, equidistant sampling requires $N_\text{fft}=\lceil2r_{max}^2/ (t_{min}\pi)\rceil$ grid points, where $\lceil \cdot\rceil$ is the ceiling  operator.
Imposing the constraint from the momentum grid one finds that the necessary grid length is $N_\text{fft}=\lceil 32k_{max}^2 t_{max}^2/ (t_{min}\pi)\rceil$.  
For each discrete time value, this implies $\mathcal{O}(N_\text{fft}\log(N_\text{fft}))$ operations compared to $\mathcal{O}(2P N_k^2)$ operations for the method presented in the previous section. 
Because we do not have to sample the fast oscillations and because the $k$ and $r$ grid are only connected by the range they cover, since $r_\text{min/max}\sim \left(k_\text{max/min}\right)^{-1}$, we can work with around 500 grid points.
For a typical calculation we use values $k_{max}\sqrt{\beta}=50,\,r_{max}/\sqrt{\beta}=40,\, t_{min}/\beta=2\times 10^{-5}$ and $t_{max}/\beta=50$ which is accurately sampled with $P=18$ as discussed in App.~\ref{app:order}. 
By comparison, a conventional FFT-based method would require around $10^9$ times more grid points and consequently $10^5$ times the number of operations.
The overhead due to the interpolations and the $N^2$ scaling of the transforms is thus vastly overcompensated by the decrease in the grid lengths.
The result is that our method allows us to compute self-energies efficiently by relying on their convolution structure, without using grids with a prohibitively large number of points.  

To further increase the efficiency of the implementation, it is useful to subtract the analytically known UV behavior, while also preventing the appearance of singularities in the IR. A detailed account of the analytical subtractions and how to apply the transformations is presented in appendix \ref{app:subtractions}.

\section{Validation}\label{sec:validation}
To test the accuracy of our method, we are now going to compare it with known analytical results, imaginary-time calculations, and state-of-the-art numerical analytic continuation.
For this comparison, we focus on the balanced case where $n_\sigma=n$ and $m_\sigma=m$ and will therefore drop the spin-dependence of all fermion propagators.

\subsection{Asymptotics}
One of the main challenges of the real-frequency formalism lies in the accurate description of the asymptotic behavior of propagators at high momenta, as well as $|\omega|\to\infty$. 
We focus on quantities that are fully determined by the effect of interactions, i.e. that vanish for the non-interacting gas, as these will provide key insights into the precision of our method. 

\begin{figure}
    \centering
    \includegraphics[width=\columnwidth]{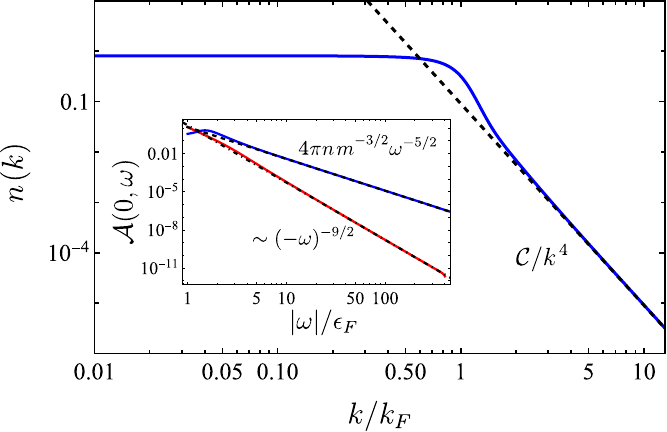}
    \caption{The fermionic spectral function $\mathcal{A}(0,\omega)$ is known to decay as $\omega^{-5/2}$ and $(-\omega)^{-9/2}$ at very high and low frequencies respectively. 
    Both limits are well reproduced by our method as is shown in the inset. 
    Furthermore, the momentum distribution at high momenta behaves as $n(k \gg k_F)\approx \mathcal{C}/k^{4}$ with $\mathcal{C}$ the Tan contact density obtained from the pair propagator.}
    \label{fig:asymptotics}
\end{figure}

At high energies, the fermionic spectral function exhibits a scattering continuum with the analytically known frequency dependence~\cite{Haussmann2009}
\begin{align}\label{eq:high_tail}
    \mathcal{A}(k,\omega\to\infty)=4\pi n m^{-3/2}\omega^{-5/2}\,,
\end{align}
where $n$ is the density of either species of fermions.
Similarly, at low frequencies, it can be shown to decay as
\begin{align}\label{eq:low_tail}
    \mathcal{A}(k,\omega\to-\infty)\sim(-\omega)^{-9/2}\,,
\end{align}
with a prefactor that has no closed analytical form. In the limit of weak coupling the latter is determined by including the behavior at high frequencies stated in Eq.~\eqref{eq:high_tail} in the evaluation of the fermonic self-energy. There one finds $4\pi n m^{-7/2} \mathcal{C}$ for the prefactor. At stronger coupling this result is modified self-consistently by a resummation to infinite order in perturbation theory.
As we show in the inset of Fig.~\ref{fig:asymptotics}, both behaviors are very well captured by our method.

The Tan contact density $\mathcal{C}$ can be calculated from the pair propagator $\mathcal{C}=i m^{2}\delta\Gamma^K(r=0,t=0)/2$ or the asymptotic decay of the momentum distribution \cite{Tan2008b} $\lim_{k\to \infty}n(k)=\mathcal{C}/k^{4}$.
For a $\Phi$-derivable theory, both calculations must coincide and therefore provide a critical test for the convergence and accuracy of our method.
In Fig.~\ref{fig:asymptotics}, the momentum distribution of the balanced, unitary Fermi gas at $T/T_F=0.2$ is compared with the asymptotic behavior $\mathcal{C}/k^{-4}$ with the contact obtained from the pair propagator. 
The excellent agreement indicates two important observations. 
On one hand, we see that the self-consistent evaluation is indeed well-converged. 
On the other hand, the clean asymptotics of $n(k)$, i.e. the absence of any numerical noise in Fig.~\ref{fig:asymptotics}, demonstrates the high fidelity, with which self-energies are evaluated, even at high momenta $k\gg k_F$ and frequencies $|\omega|\gg\epsilon_F$.

\subsection{Comparison with Matsubara formalism}
Additional insights can be gained from the comparison with results obtained in Matsubara formalism. 
Indeed, there, the quickly oscillating factor $e^{ik^2 t}$ of the real-time propagators is replaced by a Gaussian envelope $\sim e^{-k^2 \tau}$ in the imaginary time interval $\tau \in [0,\beta)$ and therefore much easier to handle. 
Consequently, the self-consistent T-matrix in Matsubara formalism is a well-established method \cite{Haussmann1994, Haussmann2007} with well-controlled errors \cite{Frank_Thesis}. 
We will leverage this to compare our new approach against this method. 
For the dimensionless Fermi energy $\epsilon_F/T$ and the contact, we find typical differences no larger than $3\%$. 
However, these integrated quantities provide little new insight over the previous tests. 
Instead, a more telling comparison can be performed between propagators directly in imaginary time.

Although the analytic continuation from Matsubara frequencies to real frequencies or the equivalent transformation from imaginary to real-time is numerically an ill-posed problem, the reverse operations are unproblematic. Imaginary time propagators are obtained from spectral functions by the generalized Laplace transform
\begin{align}
    G(k,\tau)=\int \frac{d\omega}{2\pi}\frac{e^{-\omega\tau}}{1\mp e^{-\beta\omega}} \mathcal{A}(k,\omega)\,,
\end{align}
where the upper/lower sign refers to bosons/fermions. 
It is well known that the spectrum of the latter transformation possesses only a few singular values larger than any realistic numerical resolution, which immediately explains the problems one encounters when inverting the transformation as is the case for numerical analytic continuations. 
However, it also highlights that at any finite numerical precision propagators in real frequencies contain genuinely more information than their Matsubara counterparts.
\begin{figure}
    \centering
    \includegraphics[width=\columnwidth]{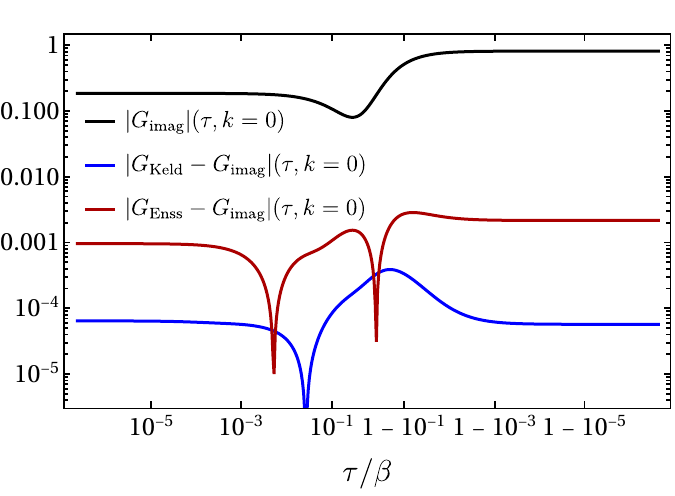}
    \caption{The absolute value of the fermionic imaginary-time propagator of the balanced unitary Fermi gas at $T/T_F=0.2$ and at vanishing momentum $k=0$ (black). It is compared with the absolute value of the difference between its evaluation directly in imaginary times and via the generalized Laplace transform of the spectral function (blue). For comparison also the results obtained with the alternative real-time approach developed in Ref.~\cite{Enss2023} is shown in red.}
    \label{fig:rel_error}
\end{figure}

\begin{figure}
    \centering
    \includegraphics[width=\columnwidth]{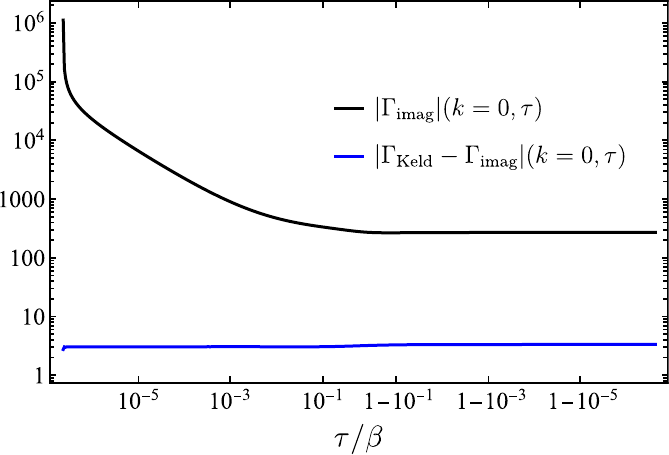}
    \caption{As for the fermionic propagator in Fig.~\ref{fig:rel_error}, the difference between the pairing propagator in imaginary times obtained from real frequencies and in Matsubara formalism (blue) is small compared to the absolute value of the paring propagator (black).}
    \label{fig:vertex_error}
\end{figure}

\begin{figure}[b]
    \centering
    \includegraphics[width=\columnwidth]{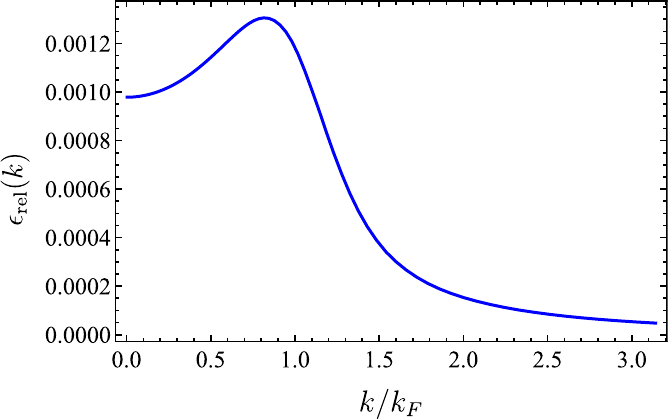}
    \caption{The integrated, relative difference between single-particle, imaginary-time propagators obtained in Matsubara and Keldysh formalism as defined in Eq.~\eqref{eq:int_error} shows very good agreement between the two independent numerical methods. The parameters are equivalent to \cref{fig:rel_error,fig:vertex_error}.}
    \label{fig:int_error}
\end{figure}

Nevertheless, the thermodynamic information contained in the imaginary time propagators needs to be fully recovered by the presented formalism. 
To this extent, both the propagators of the fermions and the pair propagator are transformed to imaginary times and compared in \cref{fig:rel_error,fig:vertex_error} with results previously obtained in the Matsubara formalism \cite{Frank2018} at zero momentum. 
Both agree for all values of $\tau/\beta$ to within $1\%$. 
For comparison, we have included the same analysis for the data obtained independently for the fermionic propagator by Enss~\cite{Enss2023}. 
The results from the method described there are found to compare well with the Matsubara results. 
The higher accuracy of the convolution-based method is probably mostly related to the longer frequency and momentum intervals accessible due to the non-equidistant sampling we use.
Furthermore, we show the integrated relative difference
\begin{align}\label{eq:int_error}
    \epsilon_\text{rel}(k)=\sqrt{\frac{\int d\tau (G_\text{imag}(k,\tau)-G_\text{Keld}(k,\tau))^2}{\int d\tau G_\text{imag}^2(k,\tau)}}
\end{align}
as a function of momentum in Fig.~\ref{fig:int_error} to highlight that this level of agreement is global and no larger differences occur at higher momenta.

We emphasize, that although at finite numerical precision, the Matsubara formalism contains less information than the method developed here, the comparison is even more critical than the previous inspection of the asymptotics.
This is because the comparison is sensitive to local quantities, like the linewidth or energy of the interacting fermions. At the precision of the numerical Matsubara calculation used in the benchmark, around 5\% of the right-singular vectors of the generalized Laplace transform can be tested. Broadly speaking, the comparison in imaginary times, therefore, tests a finite fraction of the entire information encoded in the spectrum as opposed to the single pieces of information validated by the asymptotics.

\section{Comparison with numerical analytic continuation}\label{sec:comp}
With the numerical implementation of the self-consistent T-matrix approximation validated, it is now time to turn the tables and review the accuracy of the numerical analytic continuation as the current state-of-the-art for the evaluation of spectral functions in strongly interacting systems.

\begin{figure}
    \centering
    \includegraphics[width=\columnwidth]{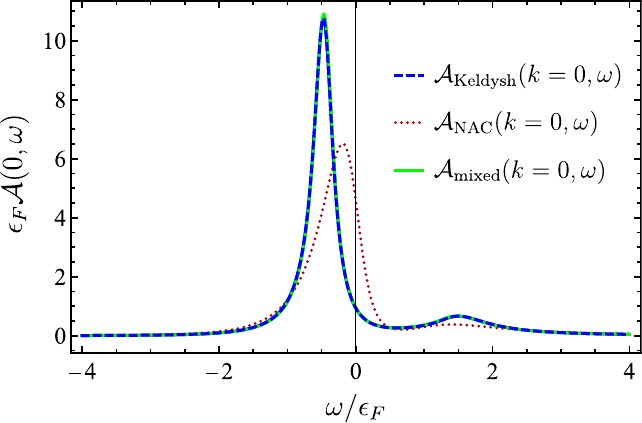}
    \caption{Comparison between single particle spectral functions at $\vect{k}=0$. $\mathcal{A}_\text{Keldysh}$ denotes the spectrum obtained by the method developed here and $\mathcal{A}_\text{NAC}$ the result from numerical analytic continuation explained in Sec.~\ref{sec:comp}. 
    The spectrum $\mathcal{A}_\text{mixed}$, which agrees almost perfectly with $\mathcal{A}_\text{Keldysh}$ is obtained by analytic continuation with $\mathcal{A}_\text{Keldysh}$ as default model, indicating that the latter is fully consistent with the imaginary time propagator used for $\mathcal{A}_\text{NAC}$.} 
    \label{fig:spectra_comp}
\end{figure}

Considering the exemplary case of the unitary Fermi gas at $T/T_F=0.2$, we calculate the Matsubara propagator in imaginary time as reported in Ref.~\cite{Frank2018}. 
We then use Bryan's algorithm \cite{Bryan1990} for the maximum entropy method \cite{Jarrell1996, Sivia2006} to perform the analytic continuation (see App.~\ref{app:NAC}). 
The maximum entropy method uses Bayesian inference to complement the limited information that can be gained from imaginary times by an entropy term $S[\mathcal{A}_0]$ relative to a default model $\mathcal{A}_0$. 
The latter is chosen as a slowly varying function with the analytically known behavior for large values of $|\omega|$ stated in \cref{eq:high_tail,eq:low_tail}.
The relative weight between the constraint in imaginary times and the entropy is a free parameter, the choice of which can be optimized in different ways, leading to somewhat different spectral functions. 
The details of this choice and the method behind the numerically continued spectral function $\mathcal{A}_\text{NAC}$ in the comparison in Fig.~\ref{fig:spectra_comp} are given in App.~\ref{app:NAC}.
As is shown in said figure for $k=0$, spectral functions obtained from numerical analytic continuation are in general too broad and typically shifted to higher frequencies, with the scattering continuum undersized when compared to $\mathcal{A}_\text{Keldysh}(k,\omega)$. 
These properties are expected since the default model has a single broad maximum. 
A pronounced gap between the two peaks of the spectrum frustrates the entropy term and is therefore underestimated with the two maxima broadened and drawn too close together by the maximum entropy method. 
We emphasize that although the qualitative features of the spectrum can be gleaned from $\mathcal{A}_\text{NAC}$, it is quantitatively a bad approximation to $\mathcal{A}_\text{Keldysh}$.

Given the large discrepancy between the two spectral functions, we also used $\mathcal{A}_\text{Keldysh}$ as the default model (i.e. $S[\mathcal{A}_\text{Keldysh}]$) to confirm that no additional error sources exist in our implementation of the maximum entropy method. 
If the default model is consistent with the Matsubara propagator in imaginary time, the Lagrange parameter $\alpha$ has no effect and the maximum entropy spectrum is identical to the default model. 
Indeed, the spectrum obtained by the same method as $\mathcal{A}_\text{NAC}$ but with $\mathcal{A}_\text{Keldysh}$ used for the default model is shown in Fig.~\ref{fig:spectra_comp} labeled $\mathcal{A}_\text{mixed}$ and agrees exceptionally well with the spectrum obtained by the newly developed method.

We conclude that although the imaginary time results obtained in Matsubara formalism are consistent with the new method, the common maximum entropy method for numerical analytic continuation is inaccurate for the strongly interacting Fermi gas at low temperatures.

\section{Spectral function near criticality}\label{sec:critical}
The large deviation between the validated method presented here, and the current standard in the form of the numerical analytic continuation, necessitates a review of previous numerical results for the spectral function of the strongly interacting Fermi gas.

\begin{figure*}[htp]
\centering
\hspace{-.25 cm}
\includegraphics[width=.34258\textwidth]{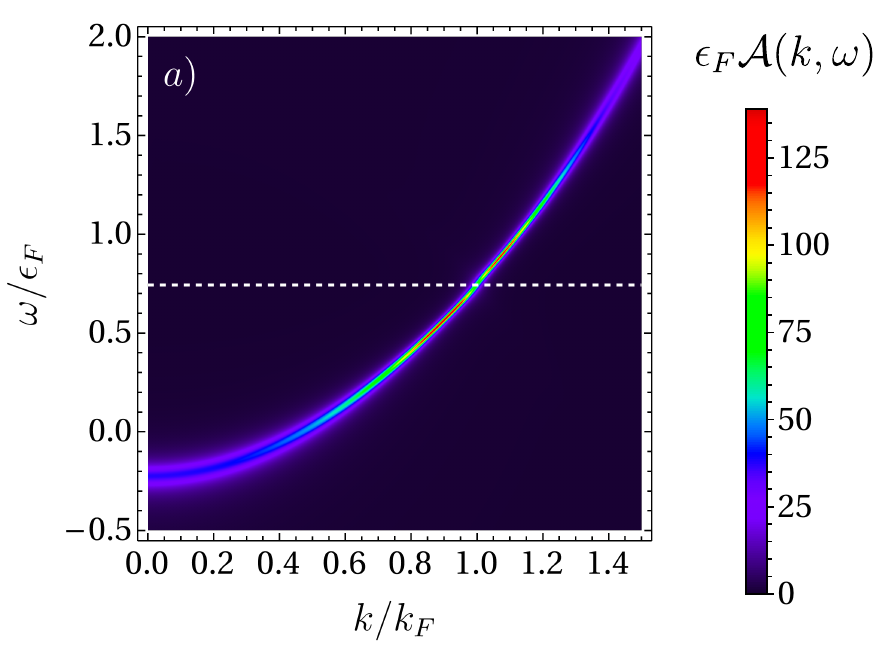}\quad
\hspace{-.4 cm}
\includegraphics[width=.330933\textwidth]{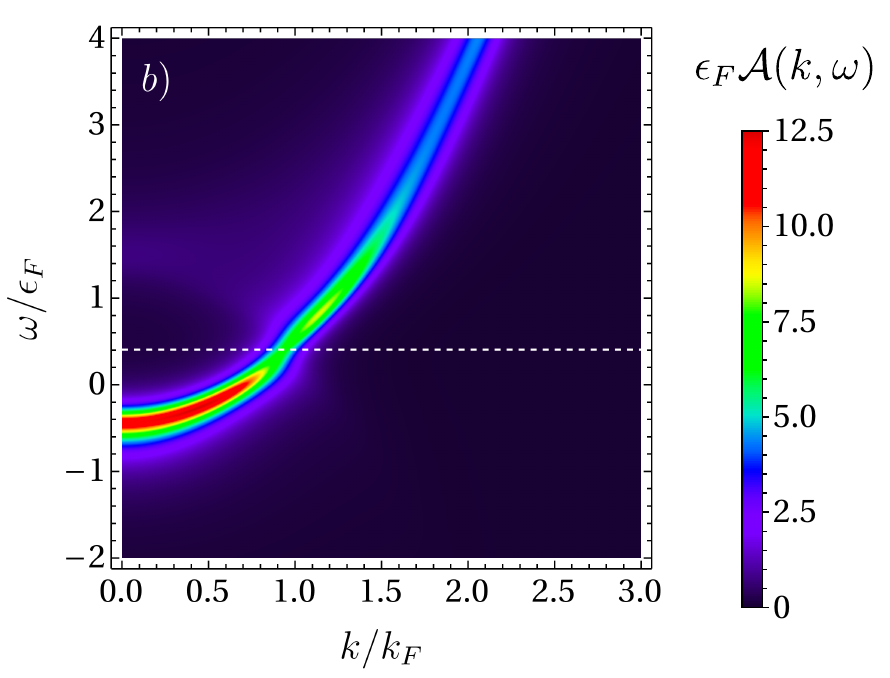}\quad
\hspace{-.4 cm}
\includegraphics[width=.330933\textwidth]{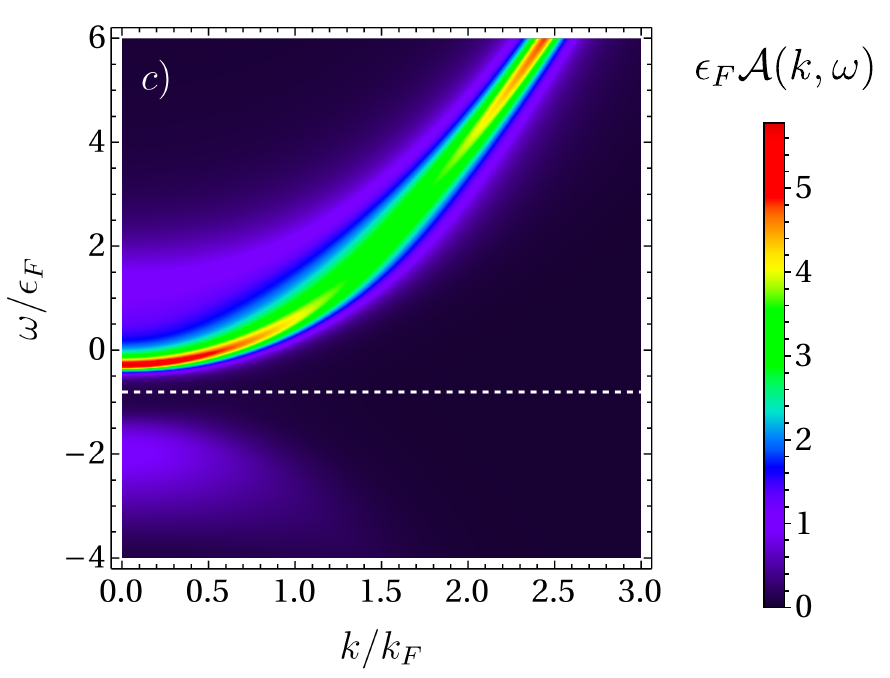}
\hspace{-.15 cm}
\caption{Spectral functions at criticality for different values of the scattering length $(k_F a)^{-1}= \{-1,0,1\}$ where the effect of the interactions increases from left to right.
The critical temperatures as determined by our methods from left to right are $T_c/T_F=\{0.068,0.156,0.205\}$ and the chemical potential $\mu/\epsilon_F=\{0.743,0.407,-0.803\}$ is indicated by the white dashed line. Both agree with the values obtained in imaginary time.}
\label{fig:contours} 
\end{figure*}

We show the results for the spectral functions at the superfluid transition at different scattering lengths in \cref{fig:contours}. We mention that the corresponding pair susceptibility $\Gamma^R(k,\omega)$ is calculated with similar accuracy as shown in App.~\ref{app:pair}.
For $k_F a=-1$ the effect of interactions is limited to a slight reduction of the spectral weight in the immediate vicinity of the Fermi surface and a broadening of the dispersion deep inside the Fermi sea. 
The latter describes the short lifetime of holes deep in the Fermi sea as no effects of artificial broadening, discussed in App.~\ref{app:fermions}, are visible in \cref{fig:contours}. 
Even at the superfluid transition of the unitary gas the dispersion in \cref{fig:contours}(b) shows only very weak signs of the nearby instability. 
There is however a weak scattering continuum visible at small momenta with its peak in reasonable agreement with the energy expected for Bogoliubov quasiparticles
\begin{align}
    E^\pm(k)=\mu\pm\sqrt{\left(\frac{k^2}{2m^*}-\mu+U\right)^2+\Delta^2}\,,
\end{align}
where the pseudogap $\Delta$, chemical potential $\mu$ and Hartree shift $U$ are fit parameters \cite{Castin2006}.

As $1/k_Fa$ is increased further, the gap in the single-particle spectrum finally opens as the low-energy effective theory transitions from interacting fermions to weakly coupled bosons formed by deeply bound pairs of fermions. 
In agreement with this picture, the hole spectrum, i.e. the occupied part of the spectrum below the chemical potential now has its maximum at $k=0$. 
Interestingly, the part of the spectrum above the chemical potential, known as the particle spectrum, at small momenta consists of a clear quasiparticle peak merged with the broad background formed by the scattering continuum. 
We emphasize that the narrow peak in the particle spectrum in \cref{fig:contours}(c) is necessitated by the large gap between the particle and hole spectrum. 
With the latter several times larger than the temperature, there are very few thermal excitations. The minimum of the upper branch in \cref{fig:contours}(c) is therefore a good approximation to the lowest state available to a single particle added to the system. Because there are no empty states with lower energy to decay to and few thermal excitations to scatter with, the state must be long-lived. 
This property is not well recovered in the analytic continuation \cite{Haussmann2009}, which otherwise captures the qualitative properties of the spectra shown in \cref{fig:contours}.

Although not visible for the small momenta shown in \cref{fig:contours}, in all cases the linewidth at high momenta decreases as $\Im{\Sigma^R(k ,\omega=k^2/(2m))}\sim k^{-1}$, in agreement with the analytic result \cite{Enss2011}. 
This reflects the vanishing role interactions play for particles with high kinetic energy exploited in our numerical implementation in Sec.~\ref{sec:method}.

The argument above regarding the lifetime of the particle spectrum illustrates the importance of an ordering principle in the energy scales. 
Especially in the highly correlated, but disordered state at temperatures above, but close to $T_c$, such simplifications are highly desirable due to the strongly interacting nature of the system. 
One therefore argues that already above the critical temperature, fermions near the Fermi surface pair up such that these pairs eventually condense at $T=T_c$. 
As breaking these pairs costs energy, one may expect a suppression of the single-particle density of states
\begin{align}
    \rho(\omega)=\int\frac{d^3 k}{(2\pi)^4}\mathcal{A}(k,\omega)
\end{align}
already above the critical temperature \cite{Norman1998}. The existence of this so-called pseudogap has been observed for example in underdoped high-$T_c$ superconductors, where its origin is, however, hard to interpret \cite{Dagotto1994} and in 2D systems \cite{Feld2011}. 
This has sparked both theoretical \cite{Randeria2010} and experimental \cite{Gaebler2010} interest in a potential pseudogap phase in the considerably simpler 3D unitary Fermi gas. 
One of the key findings obtained from the numerical analytic continuation of spectra in the self-consistent T-matrix approximation near the critical point was a much weaker pseudogap \cite{Haussmann2009} than previously reported using auxiliary field quantum Monte Carlo \cite{Magierski2009} and non-selfconsistent approximations \cite{Perali2002,Shunji2009,Chien2010,Pisani2023}. 
However, these methods either rely on numerical analytic continuation or significantly overestimate the tendency to form pairs and thus the critical temperature, which raises doubts about their validity.

\begin{figure}[b]
    \centering
    \includegraphics[width=\columnwidth]{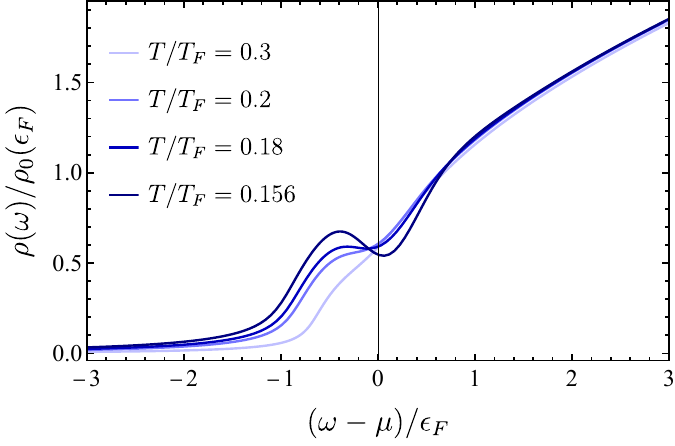}
    \caption{The single-particle density of states of the unitary Fermi gas exhibits a very weak pseudogap only for temperatures $T\lesssim 0.19\, T_F$, which is close to the critical temperature $T_c= 0.156\, T_F$. $\rho_0(\epsilon_F)= m k_F/(2\pi^2)$ denotes the density of states of the non-interacting Fermi gas.}
    \label{fig:gap}
\end{figure}

We reinvestigate the issue of the pseudogap in the normal phase of the unitary Fermi gas and report the results in Fig.~\ref{fig:gap}. 
It is found that the unitary gas develops a weak pseudogap at temperatures below $T\approx 0.19\, T_F$. 
However, even at the critical temperature $T_c= 0.156\, T_F$ the signature is not very pronounced as the abundance of thermal fluctuations broadens the spectrum, thereby precluding the possibility of a marked pseudogap. 
Our results, therefore, qualitatively confirm those previously reported using the maximum entropy method~\cite{Haussmann2009,Zwerger2014} and Pad\'e approximation \cite{Pini2019}. Moreover, the present controlled analysis settles the quantitative discrepancy between these previous results.

Although the underlying self-consistent T-matrix approximation is uncontrolled, its good agreement with experimentally observed thermodynamic properties near unitarity \cite{Haussmann2007, Frank2018, Ku2012} lends credibility also to the spectral properties reported here.

\section{Conclusions}\label{sec:Conclusion}
We have demonstrated an efficient method to calculate the spectral function of the 3D Fermi gas in the self-consistent T-matrix approximation directly in real frequencies. 
The extension to the spin- or mass-imbalanced setups is straightforward by including a spin-dependent chemical potential $\mu_\sigma$ or mass $m_\sigma$ as already shown in \cref{eq:Ham}.
The self-consistent T-matrix in imaginary time has already been applied to both mass-imbalanced \cite{Ryo2014a,Ryo2014b} and density-imbalanced systems \cite{Frank2018,Pini2023} and combinations thereof \cite{Pini2021}.
By avoiding the NAC one can for example reliably explore the predicted non-fermi liquid behaviour in the imbalanced systems, which requires accurate calculations of the linewidths.
The Keldysh formalism can also be extended to the symmetry-broken state which has already been accessed with the self-consistent T-matrix~\cite{Haussmann2007}. Furthermore, the theory can be directly applied to two dimensions. 

We emphasize that the underlying method relies only on the convolution theorem and is therefore rather versatile. 
Natural applications are Bose-Fermi models, which appear frequently in other fields of condensed matter. 
For instance, metallic quantum criticality is typically described by a Fermi surface coupled to a bosonic order parameter, which describes instabilities of the former~\cite{Loehneisen2007,LeeSS2017}. 
In two dimensions, these systems commonly show non-Fermi liquid behavior accompanied by Landau-damped critical modes, which are both primarily observed in the fermionic and bosonic spectral functions. 
Moreover, for the well-established Ising-nematic and spin-fermion models, as well as in the context of a Kondo heterostructure, it was shown that self-consistent 2PI quantum field theory reproduces imaginary-time quantum Monte Carlo data with good accuracy~\cite{Klein2020,Frank2023}. 
Since the corresponding field theory is based again on convolutions, our method, which by construction has access to both the fermionic and the bosonic spectra  (for examples cf. App.~\ref{app:pair}), seems a promising route to obtain a more quantitative understanding of the spectral functions and the identification of different scaling regimes indicated by analytic calculations~\cite{FrankPiazza2020}.

Another possible application is the quantitative analysis of quench dynamics in cold atom experiments \cite{Pruefer2018, Eigen2018, Erne2018, scazza2017repulsive}, which is well described by 2PI field theory \cite{Berges2004}. 
Notably, the method presented here is directly applicable since both the convolution structure and the conserving nature of the approximation are preserved under Wigner expansion \cite{Knoll2001}. 
Finally, the quickly growing field of pump-probe experiments in solid-state materials \cite{Basov2017} can be treated analogously \cite{Lang2023}. 
There, the order-parameter dynamics is again described by a 2PI effective field theory with dissipation. 
Our approach will be useful to develop a deeper understanding of long-lived states and scaling dynamics far from equilibrium \cite{Stojchevska2014, Liu2011, Zhang2016}.

We have presented a detailed validation of our method on the example of the Fermi gas near unitarity and outlined its applicability to a broad range of structurally similar, strongly interacting systems. 
This makes our approach a prime tool for the investigation of dynamical quantities without the need to resort to uncontrolled numerical methods.

An important future step will be the extension of the method to less restrictive approximations to the Luttinger-Ward functional, including for example higher-body correlations \cite{Vlietinck2013, Houcke2020, Milczewski2022} that lack an efficient representation in terms of convolutions. 

Furthermore, after completion of this work two other studies appeared in which the spectral functions of the strongly interacting Fermi gas are calculated in real frequencies. One using an interpolation of the self-energy \cite{Enss2023} and the other using a spectral functional approach \cite{Dizer2023}.

\section{Acknowledgements}
J.L. would like to thank Eugen Dizer for many stimulating discussions  and data comparison. We also thank Sebastian Diehl and Michele Pini for comments on the manuscript. J.L. acknowledges support by the Deutsche Forschungsgemeinschaft (DFG, German Research Foundation) CRC 1238 project number 277146847. C.H.J acknowledges support by the project No. 2021/43/P/ST2/02911 co-funded by the
National Science Centre and the European Union’s Horizon 2020 research and
innovation programme under the Marie Skłodowska-Curie grant agreement no. 945339.
B.F. acknowledges support by the DFG through the W\"urzburg-Dresden Cluster of
Excellence on Complexity and Topology in Quantum Matter -- ct.qmat (EXC 2147, project
id 390858490).
For the purpose of Open Access, the authors have applied a CC-BY public copyright licence
to any Author Accepted Manuscript (AAM) version arising from this submission. The data for all figures are available at \cite{data}. 

\appendix

\section{Details of the self-energy computation}\label{app:subtractions}

To optimize the accuracy of the loop integrals involved in the evaluation of the diagrammatic expressions that have been derived in Sec.~\ref{sec:Formalism}, it is important to ensure that the different features of terms in the self-energies are catered for. 
As we detail below, we employ several different techniques in this regard, which either allow us to reduce the domains that need to be covered numerically or improve the overall accuracy. 
Given their importance to the performance claims we make in the main text, we provide a detailed account of the essential procedures used.

\subsection{Pair propagator}\label{app:Gamma}
As stated in the main text, the contact nature of the interaction allows one to write the retarded Bethe-Salpeter equation as a Dyson equation
\begin{equation}
\Gamma^R(p)=\frac{1}{\frac{1}{g_0}-\Sigma_\Gamma^R(p)}.
\end{equation}
Including the Keldysh structure and taking advantage of the causality structure the retarded self-energy is shown in \cref{eq:PP_RetSE}.
To remove the UV divergence of $\Gamma$ arising from the contact interaction the self-energy is split into four parts
\begin{equation}
\begin{aligned}
\Sigma_\Gamma^R(p)=\Sigma_{\Gamma,v}^R(p)+\Sigma_{\Gamma,v\delta }^R(p)+\Sigma_{\Gamma, \uparrow}^R(p)+\Sigma_{\Gamma,\downarrow}^R(p).
\end{aligned}
\end{equation} 
The labeling here should be understood as the $v$ referring to a bare vacuum, $\delta v$ to a dressed vacuum and $\uparrow$ ($\downarrow$) is a contribution due to a finite density in the $\uparrow$ ($\downarrow$) species.
The bare vacuum contribution is
\begin{equation}\label{eq:Sigma_Gamma_0}
\begin{aligned}
\Sigma_{\Gamma,v}^R(p)=i\int \frac{d^4 p'}{(2\pi)^4}G^R_{\uparrow,0}(p-p')G^R_{\downarrow,0}(p').
\end{aligned}
\end{equation}
Together with the contact interaction this is exactly equal to the 2-body problem and using the Lippmann-Schwinger equation \cite{Taylor1972} one can relate this to the 2-body T-matrix
\begin{equation}\label{eq:PP_T0}
\frac{1}{g_0}-\Sigma_{\Gamma,v}^R(p)=T_0^{-1} \left(\omega-\frac{k^2}{2m_\Gamma}+\mu_\Gamma\right),
\end{equation}
where $m_\Gamma=m_\uparrow+m_\downarrow$ and $\mu_\Gamma=\mu_\uparrow+\mu_\downarrow$.
In three dimensions $T_0$ can also be computed from scattering theory as the 2-body scattering with a $\delta$-potential~\cite{Braaten2006}
\begin{equation}
     T_0^{-1}(\omega)=\frac{m_r}{\sqrt{2}\pi a_s}-\frac{m_r^{3/2}\sqrt{-\omega}}{\sqrt{2}\pi},
\end{equation}
where $m_r=m_\downarrow m_\uparrow/m_\Gamma$ is the reduced mass and $a_s$ is the scattering length.
In this way, the bare coupling $g_0$ is explicitly renormalized and connected to the experimental scattering length.
Using this procedure the full renormalized retarded pair propagator is given by
\begin{equation}\label{eq:fullGamma}
\Gamma^R(p)=\frac{1}{T_0^{-1}\left(\omega-\frac{k^2}{2m_\Gamma}+\mu_\Gamma\right)-\Sigma_{\Gamma,v\delta}^R-\Sigma_{\Gamma, \downarrow}^R-\Sigma_{\Gamma,\uparrow}^R},
\end{equation}
where the self-energy dependence on $p$ has been suppressed.
The remaining three self-energies can be read off from \cref{eq:PP_RetSE}.
One term results from modifications of the fermion propagators due to interactions and is still a vacuum term
\begin{equation}\label{PP_eq:DeltaRSEvac}
\begin{aligned}
\Sigma_{\Gamma,v\delta}^R(p)=i\int \frac{d^4 p'}{(2\pi)^4}&\Big(G^R_\uparrow(p-p')G^R_\downarrow(p')\\&-G^R_{\uparrow,0}(p-p')G^R_{\downarrow,0}(p')\Big).
\end{aligned}
\end{equation}
As opposed to \cref{eq:Sigma_Gamma_0}, the integral is finite and requires no further renormalization.
To transform this term of the self-energy one has to remove a fast oscillation with a mass given by $m_\Gamma$, such that the transformation procedure in Sec.~\ref{sec:method} is used with $\alpha=\Gamma$.

The two remaining contributions to the self-energy are caused by finite densities in either of the two species of fermions
\begin{equation}\label{eq:PP_DeltaRSEOccupation}
\begin{aligned}
\Sigma_{\Gamma,\downarrow}^R(p)&=\frac{i}{2 }\int \frac{d^4 p'}{(2\pi)^4}G^R_\uparrow(p-p')\delta G^K_\downarrow(p'),\\
\Sigma_{\Gamma,\uparrow}^R(p)&=\frac{i}{2 }\int \frac{d^4 p'}{(2\pi)^4}\delta G^K_\uparrow(p-p')G^R_\downarrow(p').
\end{aligned}
\end{equation}
As $\delta G^K_\alpha$ is cut off at large momentum by the distribution function, the self-energy $\Sigma^R_{\Gamma,\uparrow(\downarrow)}$ is transformed using the appropriate fermion mass $m_{\uparrow(\downarrow)}$. 
Next, we can use the FDR to minimize the noise of the imaginary part of $\Sigma_{\Gamma,\uparrow(\downarrow)}^R$. 
This is possible because $\delta \Sigma_\Gamma^K$ for bosons satisfies the relation
\begin{equation}\label{eq:sigKsigR_relation}
    \delta \Sigma_\Gamma^K(k,\omega) =-4 i n_B(\omega) \Im\Sigma_\Gamma^R(k,\omega).
\end{equation}
Instead of using the FDR one can also compute $\delta\Sigma_\Gamma^K$ directly using the loop integral in \cref{eq:PP_SEK}. 
This integral is well-behaved due to the fast decay of the distribution functions. 
Hence we compute $\delta\Sigma_\Gamma^K(k,\omega)$ with high accuracy and use it to remove errors in the retarded self-energy through \cref{eq:sigKsigR_relation}. 

\subsection{Fermions}\label{app:fermions}
As we have seen explicitly in the previous section, the pair propagator depends on the self-energies of the fermions, which in turn depend on the pair propagator. 
These self-consistent equations are solved iteratively, starting from the non-interacting theory. 
The fermion self-energy has several different terms with different features. 
Just as for the pair propagator it is necessary to isolate these terms and optimize the transformation of each independently. 
This is the focus of the first half of this section of the appendix. 

In the latter half, we focus on artificial broadening. 
This is only necessary to consider for the fermions as the pair propagator inherits its broadening from the fermions, whose spectra in turn will be broadened by interactions with the pairing field. 
To obtain accurate results, it is thus important to minimize the artificial broadening required for the numerical Fourier transformation of the propagators. 

Considering first the optimal transformation of the different terms in the self-energy, we start each iteration with all the propagators from the previous iteration in the $(r,t)$ basis. 
In this space, the self-energies are simple products and the retarded fermion self-energy consists of two distinct parts. 
One arises from occupation of the paring field 
\begin{equation}\label{eq:fermionSER1}
    \Sigma^R_{\sigma,1}(r,t)=\frac{i}{2}G^R_{\bar{\sigma}}(r,-t)\delta\Gamma^K(r,t),
\end{equation}
which can be transformed to momentum by removing the bare oscillations in the retarded propagator by choosing $m_\alpha=m_{\bar{\sigma}}$. 
The other contribution to the self-energy is due to occupation in the $\bar{\sigma}$-fermions
\begin{equation}
    \Sigma^R_{\sigma,2}(r,t)=\frac{i}{2}\delta G^K_{\bar{\sigma}}(r,-t)\Gamma^R(r,t).
\end{equation}
To transform this self-energy we first subtract the bare T-matrix which requires us to transform the term
\begin{equation}\label{eq:fermionSER2a}
    \Sigma^R_{\sigma,2,0}(r,t)=\frac{i}{2}\delta G^K_{\bar{\sigma}}(r,-t)T_0(r,t).
\end{equation}
This term has oscillations at large momentum set by the bare pair propagator such that we remove oscillations with a mass $m_\alpha=m_\Gamma$.
The remaining contribution 
\begin{equation}\label{eq:fermionSER2b}
    \Sigma^R_{\sigma,2,\delta}(r,t)=\frac{i}{2}\delta G^K_{\bar{\sigma}}(r,-t)\left(\Gamma^R(r,t)-T_0(r,t)\right),
\end{equation}
is generally varying significantly at short times and with a behavior not related to the bare oscillations $m_\alpha=m_\Gamma$, which dominate at larger times. 
For this reason, only the parts at times larger than a cutoff of order $(10\beta)^{-1}$ are transformed using $\alpha\rightarrow \Gamma$, while the short-time behavior is transformed without removing fast oscillations. 
This cutoff is found to be fairly robust and doesn't require fine-tuning throughout the iterations. 
By adding the three contributions in \cref{eq:fermionSER1,eq:fermionSER2a,eq:fermionSER2b} together one arrives at $\Sigma^R_{\sigma}(k,t)$.

Transforming from $(k,t)$ to $(k,\omega)$ is difficult due to a divergence at $t=0$ in $T_0$ which scales as $t^{-1/2}$. 
To remove the errors generated by this divergence, we fit $T_0(k,t=0^+)$ to $\Sigma^R_{\sigma,2,0}(k,t=0^+)$. 
By numerically transforming $\beta_k T_0(k,t)=\tilde{T}_0(k,\omega)$, with $\beta_k$ being the fitting factor, the numerical errors due to the divergence can be identified as $\tilde{T}_0(k,\omega)-\beta_k T_0(k,\omega)$. 
The procedure relies on computing $T_0(k,t)$ with very high accuracy. 
This is possible and computationally cheap because of the argument structure of $T_0(\omega-k^2/(2m_\Gamma) +\mu_\Gamma)$, as seen in \cref{eq:PP_T0}, where momenta only enter via a shift of the frequency argument. 
Furthermore, this calculation only has to be done once at initialization. 

To decrease numerical errors in the spectral functions of the fermions one can, as done in the pair propagator,  use the FDR  in \cref{eq:FDR}. 
This is done by directly computing
\begin{equation}
    \delta\Sigma_\sigma^K(r,t)=\frac{i}{2}\delta \Gamma^K(r,t)\left(\delta G^K_{\bar{\sigma}}(r,-t)-i 2\Im\Sigma_{\sigma,1}(r,t)\right).
\end{equation}
As discussed for the pair propagator, this self-energy can be computed with high precision due to the fast decay of the distribution functions.  
After subtracting the errors one arrives at the retarded self-energy for the fermions.

Having discussed how to optimally transform the different terms in fermion self-energy we now turn our attention to minimizing the effect of artificial broadening.
The artificial broadening is necessary because the bare fermion propagators exhibit a $\delta$-peak in the spectral function. 
To sample this numerical it must be made finite through artificial broadening. 
This broadening is equally important in the time domain where it ensures that the retarded propagators keep their causal structure and decay sufficiently within the length of our numerical time grid. 
Such a broadening is always introduced in the theory but should be taken to zero after the calculation is performed \cite{bruus2001}.
As this is impossible with a numerical result, it is essential to minimize the artificial broadening.
First, we discuss how the artificial broadening is included followed by how it is minimized throughout the iterations. 

Conventionally, this broadening is included as a constant and positive imaginary contribution to the bare retarded propagator as 
\begin{equation}\label{eq:eta_std}
    \tilde{G}_{0,\eta}^R(\omega)=\frac{1}{\omega+i \eta},
\end{equation}
where the propagator has been written on the transformed frequency grid discussed in Sec.~\ref{sec:method}.
The corresponding Lorentzian spectral function decays as $\omega^{-2}$. 
If such tails are included in the calculation, then they will have a large impact on the high frequency tails of the spectral functions. 
Through the convolutions, this then leads to significant errors, for example on the Tan contact density extracted from the pair-propagator as well as total fermion density.
It is therefore essential to avoid these tails as much as possible. 
To this extent, we add the finite broadening through an artificial frequency-dependent retarded self-energy of the form
\begin{equation}\label{eq:NM_Sigr0}
    \Sigma_\rho^R(\omega)=\frac{1}{\rho \omega - \frac{1}{\omega+i}},
\end{equation}
where $\rho$ is a positive computational parameter.
The broadened bare propagator is then given by
\begin{equation}\label{eq:NM_Gr0}
    G_{0,\eta}^R(\omega)=\frac{1}{\omega-\eta\Sigma^R_\rho(\omega)}.
\end{equation}
Because the artificial self-energy is retarded, it preserves the causality of the propagator \cite{kamenev2011}.
Differently from the standard case in \cref{eq:eta_std}, our artificial broadening also contains a real part, which is necessary to satisfy the Kramers-Kronig relations \cite{morawetz2018}. However, if one chooses the two artificial parameters such that  $\rho\gg 1\gg \eta$ and $\eta \rho\ll1$ then the resulting spectrum is very well described as
\begin{equation}
    A_{0,\eta}(\omega)=\frac{2\eta}{\eta^2+\omega^2+\rho^2\left(\omega^4+\omega^6\right)}.
\end{equation}
In practice, we satisfy the limit by choosing $\rho=10$ and $\eta\leq 10^{-2}$.
With this choice the spectral function close to the bare dispersion (meaning $\omega=0$), is almost Lorentzian but as one deviates from the bare dispersion the bare spectral function now decays as $\omega^{-6}$.
One can find analytic expressions for the poles of \cref{eq:NM_Gr0}, which allows us to compute the Fourier transforms to $(r,t)$ analytically, ensuring no additional errors are added in our subtraction schemes.
Including the additional broadening in this way makes it possible to keep the artificial width $\eta$, close to the dispersion, large enough to allow for sampling of the peak without introducing significant errors in the frequency tails of the spectral function. 

Throughout the self-consistent iterations, the fermions acquire a natural broadening due to the interactions.  
The result is that less artificial broadening is needed to reliably perform the transformations and it is thus preferable to remove all the superfluous artificial broadening at each iteration. 
For this procedure, we first compute the broadening of the quasiparticle pole for each momentum
\begin{equation}
    \eta_0(k)=-\Im\Sigma^R_\sigma(k,\epsilon_0(k)),
\end{equation}
where $\epsilon_0(k)$ satisfies 
\begin{equation}
\left(G^R_{0,\sigma}(k,\epsilon_0(k))\right)^{-1}-\Re\Sigma^R_\sigma(k,\epsilon_0(k))=0.
\end{equation}
Whenever $\eta_0$ is greater than $\eta$ no artificial broadening has to be added.
In the case where $\eta_0<\eta$ one can identify the minimal artificial broadening needed to ensure we can resolve the sharp features in the spectrum.
This minimal broadening $\eta_m$ is defined as
\begin{equation}
    \eta_m(k)=\theta\big(\eta-\eta_0(k)\big)(\eta-\eta_0(k)).
\end{equation}
We then compute the smallest imaginary value, $\eta_s$, over all $k$ and $\omega$, of the self-energy when the minimum quasiparticle linewidth has been renormalized
\begin{equation}
    \eta_s=2\,\text{Min}\,\abs{\Im\left(\Sigma^R_\sigma(k,\omega)+\eta_m(k)\Sigma_\rho^R\big(k^2 v_\alpha +\mu_\alpha+\omega\big)\right)},
\end{equation}
where $\Sigma^R_\rho$ is defined in \cref{eq:NM_Sigr0}.
By removing this constant and adding $\eta_m(k)$ the maximum amount of artificial broadening has been removed while retaining the causal structure of self-energy at all points in $p$-space.
This procedure gives a quantitative improvement but is not essential for the stability of the method. Typical values of $\eta_s$ are on the order of $10^{-3}/\beta$.
The final retarded self-energy is then
\begin{equation}
    \Sigma^R_\sigma(k,\omega)\rightarrow\Sigma^R_{\sigma}(k,\omega)+i \eta_s+\eta_m \Sigma_\rho^R(k,\omega),
\end{equation}
and the retarded fermion propagator can be computed through the Dyson equation \cref{eq:RetDyson}. 
With this procedure, we have ensured that at each iteration the smallest possible artificial broadening is added. 

For imbalanced systems, one repeats all the same steps for the second fermion species $\bar{\sigma}$.   

\section{Interpolation order}\label{app:order}
The upper error bound between a function $f(x)$ and its $q$-point Hermite interpolation of order $P$ at the point $x$, $h(x)$, is given by \cite{Phillips2003}
\begin{equation}
\abs{f(x)-h(x)}\leq \abs{\frac{f^{(qP)}(a)}{q P!}\prod_{i=1}^{q}\left(x-x_i\right)^P},  
\end{equation}
where $a\in\{x_1,x_{q}\}$ is the value that maximizes $f^{(qP)}(a)$.
Specifying this for the 2-point case where $q=2$ and $f(x)=\sin(x b)/b$ 
\begin{equation}
\begin{aligned}
 \max\abs{\frac{\sin(x b)}{b}-h(x)}=&\max\Bigg\vert\left(\frac{\dd^{2P}}{\dd y^{2P}}\frac{ \sin(y b )}{b}\right)_{y=a}\frac{1}{(2P)!}\\&\times\left(x-x_n\right)^P\left(x-x_{n+1}\right)^P\Bigg\vert\\<&\max\left[\frac{b^{2P-1}}{(2P)!}\left(\frac{\Delta x_N }{2}\right)^{2P}\right],
\end{aligned}
\end{equation}
where $\Delta x_N=x_{N+1}-x_N$ and it has been assumed that the largest grid spacing is at the end of the grid. 
As the backward and forward transformations require two different interpolations one can put an upper limit on the interpolation error given by 
\begin{equation}
    \epsilon = \max\left[\frac{r_N^{2P-1}}{(2P)!}\left(\frac{\Delta k_N }{2}\right)^{2P},\frac{k_N^{2P-1}}{(2P)!}\left(\frac{\Delta r_N }{2}\right)^{2P}\right].
\end{equation}
From this relation, one can choose the appropriate interpolation order. 
This error gives an upper bound and therefore generally gives a larger than necessary $P$. 

\section{Maximum Entropy method}\label{app:NAC}
In this appendix, we provide the details of the maximum entropy method used for the numerical analytic continuation in Sec.~\ref{sec:comp}.

As explained in the main text, numerical propagators with a finite accuracy in real frequencies contain significantly more information than their counterparts in imaginary time or Matsubara frequency. Specifically, NAC aims to minimize the dimensionless distance from the Matsubara propagator in imaginary times
\begin{align}\label{eq:chi}
    |\chi(k)|^2=\sum_i\left(G(k,\tau_i)-\int d\omega \frac{e^{-\tau_i \omega}}{1+e^{-\beta \omega}}\mathcal{A}(k,\omega)\right)^2\,,
\end{align} 
where in our case the imaginary times $\tau_i$ are logarithmically spaced near 0 and $\beta$. 
The local relative numerical accuracy of the Matsubara propagator is $\sim 10^{-8}$. 
Consequently, upon discretization, only a few of the right-singular vectors of the integral transformation are indeed fixed by the minimization of $|\chi(k)|^2\lesssim 10^{-8}$. 
Using an optimized grid with 400 points that take into account the known asymptotic decay of $\mathcal{A}$ at small and high frequencies and using cubic splines for the integration, only around 80 of the right-singular vectors of the integral transformation in \eqref{eq:chi} provide meaningful information. 
Consequently, at the given accuracy an infinite set of spectral functions $\mathcal{A}$ produce the same imaginary-time propagator.

To resolve this ambivalence one needs to introduce some prior information. Using Bayesian statistics \cite{Jarrell1996,Sivia2006} one argues that instead of minimizing $|\chi(k)|^2$ the spectral function should maximize
\begin{align}
    Q(k)=\alpha S(k)-\frac{1}{2}|\chi(k)|^2
\end{align}
where
\begin{align}
\begin{split}
    S[\mathcal{A}_0](k)=&\int d\omega\left(\mathcal{A}(k,\omega)-\mathcal{A}_0(k,\omega)\right.\\&\left.- \mathcal{A}(k,\omega)\ln{(\mathcal{A}(k,\omega)/\mathcal{A}_0(k,\omega))}\right)
\end{split}
\end{align}
is the entropy relative to a smooth default model $\mathcal{A}_0$ containing analytically known properties. Since $|\chi(k)|^2$ provides little information in the high and low-frequency tails, we use the default model
\begin{align}
    \mathcal{A}_0(k,\omega-k^2/(2m))=\frac{\gamma\left(\sqrt{\frac{\delta}{\gamma}+\omega^2}+\omega
   \right)}{\left(\rho +\omega^2\right)^{7/4}}\,,
\end{align}
with $\gamma=2\pi n m^{-3/2}$ and 
$\delta=8\pi n m^{-7/2} \mathcal{C}$, which satisfies the known asymptotic behavior from Eq.~\eqref{eq:high_tail} for $\omega \to \infty$ and from Eq.~\eqref{eq:low_tail} for $\omega \to -\infty$. Furthermore, $\rho$ is determined by the normalization condition $\int_\omega\mathcal{A}_0(k,\omega)=1$ and on the typical energy scales set by $T$ and $\epsilon_F$ the default model is a smooth, slowly varying function.

\begin{figure}
    \centering
    \includegraphics[width=\columnwidth]{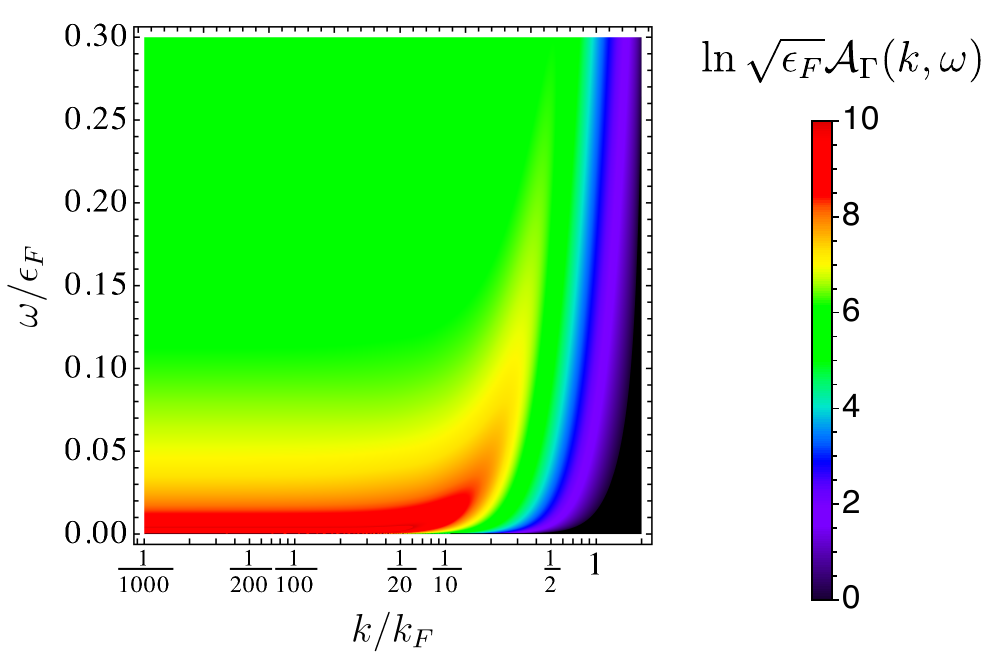}
    \caption{Spectral function of the pair susceptibility for the unitary balanced system at the critical temperature $T/T_F=0.156$. The logarithmic color scale highlights structures beyond the singularity at $(k=0,\omega=0)$.}
    \label{fig:pairSpectrum}
\end{figure}
Due to the large information gap left by $|\chi(k)|^2$ the choice of the Lagrange parameter $\alpha$ is very important. We follow Bryan's method \cite{Bryan1990} and obtain the optimal spectral function $\mathcal{A}_\alpha(k)$ for each value of $\alpha$. These are then averaged over $\alpha$ to obtain the best estimate for the spectral function
\begin{align}
    \mathcal{A}_\text{NAC}(k)=\mathcal{N}\!\int\! d\alpha \mathcal{A}_\alpha(k) f(\alpha) e^{Q(\mathcal{A}_\alpha(k))}\prod_i\left(\frac{\alpha}{\alpha+\lambda_i}\right)^{\!1/2}\,.
\end{align}
Here $\mathcal{N}$ denotes the normalization and $\lambda_i$ are the eigenvalues of the curvature matrix of $|\chi(k)|^2$ in the space of spectral functions. 
Finally, $f(\alpha)$ is an undetermined, dimensionless scaling function, for which we choose $f(\alpha)=\alpha^{\delta}$. 
This choice determines the relative importance of entropy and constraints from imaginary times. 
For too large values of $\delta$ the entropy is weighted too heavily such that eventually $\mathcal{A}_\text{NAC}(\omega,k)\underset{\delta\to\infty}{\longrightarrow}\mathcal{A}_0(\omega,k)$. 
For too small values of $\delta$ on the other hand, $\mathcal{A}_\text{NAC}(\omega,k)$ develops oscillatory instabilities. 
Hence, the choice of the scaling function $f(\alpha)$ has a significant impact on the spectral function $\mathcal{A}_\text{NAC}$. 
We fix $\delta=-1.74$, which is the smallest value, that induces no oscillations in the asymptotic behavior at large positive or negative frequencies. 
Since this condition is somewhat arbitrary, we emphasize that the qualitative difference with the spectral function obtained directly in real frequencies $\mathcal{A}_\text{Keldysh}(\omega,k)$, which we report in Fig.~\ref{fig:spectra_comp}, remains for all stable values of $\delta$.

\section{Pair susceptibility}\label{app:pair}
A recent theoretical proposal has suggested that it might be possible to probe the pair propagator using Raman spectroscopy techniques \cite{diessel2022gMol}. 
Here, the pair propagator for the imbalanced system was argued to contain information about the FFLO phase transition giving it experimental relevance. 
Another example where the spectrum of the pair propagator can be of use is in questions regarding transport \cite{Frank2020}.
Here computation of dynamical transport coefficients, like the dynamical shear viscosity, is simplified if one has access to frequency-resolved $\Gamma^R(k,\omega)$. 
For bosonic propagators, the NAC is even more challenging \cite{Nogaki2023}, making reliable results difficult to acquire.
With our method the full frequency dependence of $\Gamma^R$ is computed at no additional cost and bosonic spectra are in general computed with the same high accuracy as their fermionic counterparts. 
To highlight this we plot, in \cref{fig:pairSpectrum}, the spectral function of the pair-propagator for the balanced case at unitarity and at the critical temperature. 
Here the singularity at $\omega=0$ for small momentum and the slowly decaying frequency tails observed in imaginary time in \cref{fig:vertex_error}, also appear in real frequencies. 

\bibliography{3DFG_biblio}

@article{Timmermans2001,
title = {Prospect of creating a composite Fermi–Bose superfluid},
journal = {Physics Letters A},
volume = {285},
number = {3},
pages = {228-233},
year = {2001},
issn = {0375-9601},
doi = {https://doi.org/10.1016/S0375-9601(01)00346-2},
url = {https://www.sciencedirect.com/science/article/pii/S0375960101003462},
author = {Eddy Timmermans and Kyoko Furuya and Peter W. Milonni and Arthur K. Kerman}
}

@article{Holland2001,
  title = {Resonance Superfluidity in a Quantum Degenerate Fermi Gas},
  author = {Holland, M. and Kokkelmans, S. J. J. M. F. and Chiofalo, M. L. and Walser, R.},
  journal = {Phys. Rev. Lett.},
  volume = {87},
  issue = {12},
  pages = {120406},
  numpages = {4},
  year = {2001},
  month = {Aug},
  publisher = {American Physical Society},
  doi = {10.1103/PhysRevLett.87.120406},
  url = {https://link.aps.org/doi/10.1103/PhysRevLett.87.120406}
}

@article{Schmidt2012,
	author = {{Schmidt, R.} and {Rath, S.P.} and {Zwerger, W.}},
	title = {Efimov physics beyond universality},
	DOI= "10.1140/epjb/e2012-30841-3",
	url= "https://doi.org/10.1140/epjb/e2012-30841-3",
	journal = {Eur. Phys. J. B},
	year = 2012,
	volume = 85,
	number = 11,
	pages = "386",
	month = "",
}

@article{Tan2008c,
title = {Generalized virial theorem and pressure relation for a strongly correlated Fermi gas},
journal = {Annals of Physics},
volume = {323},
number = {12},
pages = {2987-2990},
year = {2008},
issn = {0003-4916},
doi = {https://doi.org/10.1016/j.aop.2008.03.003},
url = {https://www.sciencedirect.com/science/article/pii/S0003491608000420},
author = {Shina Tan}
}

@article{Tan2008b,
title = {Energetics of a strongly correlated Fermi gas},
journal = {Annals of Physics},
volume = {323},
number = {12},
pages = {2952-2970},
year = {2008},
issn = {0003-4916},
doi = {https://doi.org/10.1016/j.aop.2008.03.004},
url = {https://www.sciencedirect.com/science/article/pii/S0003491608000456},
author = {Shina Tan}
}

@article{Tan2008a,
title = {Large momentum part of a strongly correlated Fermi gas},
journal = {Annals of Physics},
volume = {323},
number = {12},
pages = {2971-2986},
year = {2008},
issn = {0003-4916},
doi = {https://doi.org/10.1016/j.aop.2008.03.005},
url = {https://www.sciencedirect.com/science/article/pii/S0003491608000432},
author = {Shina Tan}
}

@article{Berges2004,
    author = {Berges, Jürgen},
    title = "{Introduction to Nonequilibrium Quantum Field Theory}",
    journal = {AIP Conference Proceedings},
    volume = {739},
    number = {1},
    pages = {3-62},
    year = {2004},
    month = {12},
    issn = {0094-243X},
    doi = {10.1063/1.1843591},
    url = {https://doi.org/10.1063/1.1843591},
    eprint = {https://pubs.aip.org/aip/acp/article-pdf/739/1/3/11882676/3\_1\_online.pdf},
}

@article{Knoll2001,
title = {Exact Conservation Laws of the Gradient Expanded Kadanoff–Baym Equations},
journal = {Annals of Physics},
volume = {293},
number = {2},
pages = {126-146},
year = {2001},
issn = {0003-4916},
doi = {https://doi.org/10.1006/aphy.2001.6185},
url = {https://www.sciencedirect.com/science/article/pii/S0003491601961858},
author = {J. Knoll and Yu.B. Ivanov and D.N. Voskresensky}
}

@Article{Basov2017,
author={Basov, D. N.
and Averitt, R. D.
and Hsieh, D.},
title={Towards properties on demand in quantum materials},
journal={Nature Materials},
year={2017},
month={Nov},
day={01},
volume={16},
number={11},
pages={1077-1088},
issn={1476-4660},
doi={10.1038/nmat5017},
url={https://doi.org/10.1038/nmat5017}
}

@article{Liu2011,
  title = {Photoinduced Phase Transitions by Time-Resolved Far-Infrared Spectroscopy in ${\mathrm{V}}_{2}{\mathrm{O}}_{3}$},
  author = {Liu, M. K. and Pardo, B. and Zhang, J. and Qazilbash, M. M. and Yun, Sun Jin and Fei, Z. and Shin, Jun-Hwan and Kim, Hyun-Tak and Basov, D. N. and Averitt, R. D.},
  journal = {Phys. Rev. Lett.},
  volume = {107},
  issue = {6},
  pages = {066403},
  numpages = {5},
  year = {2011},
  month = {Aug},
  publisher = {American Physical Society},
  doi = {10.1103/PhysRevLett.107.066403},
  url = {https://link.aps.org/doi/10.1103/PhysRevLett.107.066403}
}

@article{Stojchevska2014,
author = {L. Stojchevska  and I. Vaskivskyi  and T. Mertelj  and P. Kusar  and D. Svetin  and S. Brazovskii  and D. Mihailovic },
title = {Ultrafast Switching to a Stable Hidden Quantum State in an Electronic Crystal},
journal = {Science},
volume = {344},
number = {6180},
pages = {177-180},
year = {2014},
doi = {10.1126/science.1241591},
URL = {https://www.science.org/doi/abs/10.1126/science.1241591},
eprint = {https://www.science.org/doi/pdf/10.1126/science.1241591}}

@Article{Zhang2016,
author={Zhang, Jingdi
and Tan, Xuelian
and Liu, Mengkun
and Teitelbaum, S. W.
and Post, K. W.
and Jin, Feng
and Nelson, K. A.
and Basov, D. N.
and Wu, Wenbin
and Averitt, R. D.},
title={Cooperative photoinduced metastable phase control in strained manganite films},
journal={Nature Materials},
year={2016},
month={Sep},
day={01},
volume={15},
number={9},
pages={956-960},
issn={1476-4660},
doi={10.1038/nmat4695},
url={https://doi.org/10.1038/nmat4695}
}

@book{Jauho2007, title={Quantum Kinetics in Transport and Optics of Semiconductors}, publisher={Springer}, author={Haug, Hartmut and Jauho, Antti-Pekka}, year={2007}}

@book{Taylor1972, title={Scattering Theory}, publisher={Wiley}, author={John R. Taylor}, year={1972}}

@book{bruus2001, title={Many-Body quantum theory in condensed matter physics}, publisher={Oxford University Press}, author={Bruus,  Henrik and Flensberg, Karsten}, year={2001}}

@book{altland2010, place={Cambridge}, edition={2}, title={Condensed Matter Field Theory}, DOI={10.1017/CBO9780511789984}, publisher={Cambridge University Press}, author={Altland, Alexander and Simons, Ben D.}, year={2010}}

@book{kamenev2011, place={Cambridge}, title={Field Theory of Non-Equilibrium Systems}, DOI={10.1017/CBO9781139003667}, publisher={Cambridge University Press}, author={Kamenev, Alex}, year={2011}}

@book{morawetz2018, title={Interacting Systems far from Equilibrium}, DOI={10.1093/oso/9780198797241.001}, publisher={Oxford University Press}, author={Morawetz, Klaus}, year={2018}}

@Book{Zwerger2012,
  Title                    = {{T}he {BCS}-{BEC} {C}rossover and the {U}nitary {F}ermi {G}as},
  Editor                   = {Zwerger, W.},
  Publisher                = {Lecture Notes in Physics, Vol. 836},
  Year                     = {2012},

  Address                  = {Springer, Berlin, Heidelberg}
}

@article{Flambaum1993,
  title = {Calculation of the scattering length in atomic collisions using the semiclassical approximation},
  author = {Gribakin, G. F. and Flambaum, V. V.},
  journal = {Phys. Rev. A},
  volume = {48},
  issue = {1},
  pages = {546--553},
  numpages = {0},
  year = {1993},
  month = {Jul},
  publisher = {American Physical Society},
  doi = {10.1103/PhysRevA.48.546},
  url = {https://link.aps.org/doi/10.1103/PhysRevA.48.546}
}

@article{Harabati1999,
  title = {Analytical calculation of cold-atom scattering},
  author = {Flambaum, V. V. and Gribakin, G. F. and Harabati, C.},
  journal = {Phys. Rev. A},
  volume = {59},
  issue = {3},
  pages = {1998--2005},
  numpages = {0},
  year = {1999},
  month = {Mar},
  publisher = {American Physical Society},
  doi = {10.1103/PhysRevA.59.1998},
  url = {https://link.aps.org/doi/10.1103/PhysRevA.59.1998}
}

@article{Schmidt2013,
  title = {Field-theoretical study of the Bose polaron},
  author = {Rath, Steffen Patrick and Schmidt, Richard},
  journal = {Phys. Rev. A},
  volume = {88},
  issue = {5},
  pages = {053632},
  numpages = {16},
  year = {2013},
  month = {Nov},
  publisher = {American Physical Society},
  doi = {10.1103/PhysRevA.88.053632},
  url = {https://link.aps.org/doi/10.1103/PhysRevA.88.053632}
}

@book{Gottfried2003,
  title={Quantum Mechanics: Fundamentals},
  author={Gottfried, K. and Yan, T.M.},
  isbn={9780387955766},
  lccn={2002030571},
  series={Graduate Texts in Contemporary Physics},
  url={https://books.google.de/books?id=8gFX-9YcvIYC},
  year={2003},
  publisher={Springer New York}
}

@article{RammerRev,
  title = {Quantum field-theoretical methods in transport theory of metals},
  author = {Rammer, J. and Smith, H.},
  journal = {Rev. Mod. Phys.},
  volume = {58},
  issue = {2},
  pages = {323--359},
  numpages = {0},
  year = {1986},
  month = {Apr},
  publisher = {American Physical Society},
  doi = {10.1103/RevModPhys.58.323},
  url = {https://link.aps.org/doi/10.1103/RevModPhys.58.323}
}

@article{Chin2010,
  title = {Feshbach resonances in ultracold gases},
  author = {Chin, Cheng and Grimm, Rudolf and Julienne, Paul and Tiesinga, Eite},
  journal = {Rev. Mod. Phys.},
  volume = {82},
  issue = {2},
  pages = {1225--1286},
  numpages = {0},
  year = {2010},
  month = {Apr},
  publisher = {American Physical Society},
  doi = {10.1103/RevModPhys.82.1225},
  url = {https://link.aps.org/doi/10.1103/RevModPhys.82.1225}
}

@InCollection{Zwerger2014,
  Title                    = {{S}trongly interacting {F}ermi gases},
  Author                   = {Zwerger, W.},
  Booktitle                = {Quantum matter at ultralow temperatures, Proceedings of the International School of Physics 'Enrico {F}ermi', Course 191, Varenna, 7-15 July 2014},
  Publisher                = {IOS Press},
  Year                     = {2016},

  Address                  = {Amsterdam},
  Editor                   = {Inguscio, M. and Ketterle, W. and Stringari, S. and Roati, G.},
  Pages                    = {63-141}
}

@InCollection{Castin2006,
  title={{B}asic theory tools for degenerate Fermi gases},
  author={Castin, Y.},
  Booktitle={{U}ltra-cold {F}ermi {G}ases, Proceedings of the International School of Physics 'Enrico {F}ermi', Course 164, Varenna, 20-30 June 2006},
  Editor={Inguscio, M. and Ketterle, W. and Salomon, C.},
  year={2008},
  publisher={IOS Press},
  address={Amsterdam},
  pages={289-349}
}

@Article{Feshbach1958,
  Title                    = {{A} {U}nified {T}heory of {N}uclear {R}eactions},
  Author                   = {Feshbach, H.},
  Journal                  = {Annals of Physics},
  Year                     = {1958},
  Pages                    = {357-390},
  Volume                   = {5}
}

@Article{Fano1961,
  Title                    = {{E}ffects of {C}onfiguration {I}nteraction on {I}ntensities and {P}hase {S}hifts},
  Author                   = {Fano, U.},
  Journal                  = {Phys. Rev.},
  Year                     = {1961},
  Number                   = {6},
  Pages                    = {1866},
  Volume                   = {124}
}

@article{Nozieres1985,
author={Nozi{\`e}res, P.
and Schmitt-Rink, S.},
title={Bose condensation in an attractive fermion gas: From weak to strong coupling superconductivity},
journal={Journal of Low Temperature Physics},
year={1985},
month={May},
day={01},
volume={59},
number={3},
pages={195-211},
issn={1573-7357},
doi={10.1007/BF00683774},
url={https://doi.org/10.1007/BF00683774}
}

@article{Veillette2007,
  title = {Large-$N$ expansion for unitary superfluid Fermi gases},
  author = {Veillette, Martin Y. and Sheehy, Daniel E. and Radzihovsky, Leo},
  journal = {Phys. Rev. A},
  volume = {75},
  issue = {4},
  pages = {043614},
  numpages = {13},
  year = {2007},
  month = {Apr},
  publisher = {American Physical Society},
  doi = {10.1103/PhysRevA.75.043614},
  url = {https://link.aps.org/doi/10.1103/PhysRevA.75.043614}
}

@Article{Melo1993,
  Title                    = {{C}rossover from {BCS} to {B}ose {S}uperconductivity: {T}ransition {T}emperature and {T}ime-{D}ependent {Ginzburg-Landau} {T}heory},
  Author                   = {{Sa de Melo}, C. A. R. and Randeria, Mohit and Engelbrecht, Jan R.},
  Journal                  = {Phys. Rev. Lett.},
  Year                     = {1993},
  Number                   = {19},
  Pages                    = {3202},
  Volume                   = {71}
}

@Article{Chevy2006,
  Title                    = {{U}niversal phase diagram of a strongly interacting {F}ermi gas with unbalanced spin populations},
  Author                   = {Chevy, F.},
  Journal                  = {Phys. Rev. A},
  Year                     = {2006},
  Number                   = {6},
  Pages                    = {063628},
  Volume                   = {74}
}

@article{Hofmann2010,
  title = {Quantum Anomaly, Universal Relations, and Breathing Mode of a Two-Dimensional Fermi Gas},
  author = {Hofmann, Johannes},
  journal = {Phys. Rev. Lett.},
  volume = {108},
  issue = {18},
  pages = {185303},
  numpages = {5},
  year = {2012},
  month = {May},
  publisher = {American Physical Society},
  doi = {10.1103/PhysRevLett.108.185303},
  url = {https://link.aps.org/doi/10.1103/PhysRevLett.108.185303}
}

@article{Gao2012,
  title = {Breathing mode of two-dimensional atomic Fermi gases in harmonic traps},
  author = {Gao, Chao and Yu, Zhenhua},
  journal = {Phys. Rev. A},
  volume = {86},
  issue = {4},
  pages = {043609},
  numpages = {4},
  year = {2012},
  month = {Oct},
  publisher = {American Physical Society},
  doi = {10.1103/PhysRevA.86.043609},
  url = {https://link.aps.org/doi/10.1103/PhysRevA.86.043609}
}

@article{Enss2011,
title = {Viscosity and scale invariance in the unitary Fermi gas},
journal = {Annals of Physics},
volume = {326},
number = {3},
pages = {770-796},
year = {2011},
issn = {0003-4916},
doi = {https://doi.org/10.1016/j.aop.2010.10.002},
url = {https://www.sciencedirect.com/science/article/pii/S000349161000179X},
author = {Tilman Enss and Rudolf Haussmann and Wilhelm Zwerger}
}

@article{Frank2018,
author={Frank, B.
and Lang, J.
and Zwerger, W.},
title={Universal Phase Diagram and Scaling Functions of Imbalanced Fermi Gases},
journal={Journal of Experimental and Theoretical Physics},
year={2018},
month={Nov},
day={01},
volume={127},
number={5},
pages={812-825},
issn={1090-6509},
doi={10.1134/S1063776118110031},
url={https://doi.org/10.1134/S1063776118110031}
}

@phdthesis{Frank_Thesis,
	author = {Frank, Bernhard},
	title = {Thermodynamics and Transport in Fermi Gases near Unitarity},
	year = {2019},
	school = {Technische Universität München},
	pages = {265},
	note = {},
}

@article{Frank2020,
  title = {Quantum critical thermal transport in the unitary Fermi gas},
  author = {Frank, Bernhard and Zwerger, Wilhelm and Enss, Tilman},
  journal = {Phys. Rev. Res.},
  volume = {2},
  issue = {2},
  pages = {023301},
  numpages = {15},
  year = {2020},
  month = {Jun},
  publisher = {American Physical Society},
  doi = {10.1103/PhysRevResearch.2.023301},
  url = {https://link.aps.org/doi/10.1103/PhysRevResearch.2.023301}
}

@misc{Lang2023,
      title={Field theory for the dynamics of the open $O(N)$ model}, 
      author={Johannes Lang and Michael Buchhold and Sebastian Diehl},
      year={2023},
      eprint={2310.06892},
      archivePrefix={arXiv},
      primaryClass={cond-mat.stat-mech}
}

@misc{Enss2023,
      title={Particle and pair spectra for strongly correlated Fermi gases: a real-frequency solver}, 
      author={Tilman Enss},
      year={2023},
      eprint={2311.05443},
      archivePrefix={arXiv},
      primaryClass={cond-mat.quant-gas}
}

@misc{Dizer2023,
      title={Spectral properties and observables in ultracold Fermi gases}, 
      author={Eugen Dizer and Jan Horak and Jan M. Pawlowski},
      year={2023},
      eprint={2311.16788},
      archivePrefix={arXiv},
      primaryClass={cond-mat.quant-gas}
}

@article{Haussmann1994,
  title = {Properties of a Fermi liquid at the superfluid transition in the crossover region between BCS superconductivity and Bose-Einstein condensation},
  author = {Haussmann, R.},
  journal = {Phys. Rev. B},
  volume = {49},
  issue = {18},
  pages = {12975--12983},
  numpages = {0},
  year = {1994},
  month = {May},
  publisher = {American Physical Society},
  doi = {10.1103/PhysRevB.49.12975},
  url = {https://link.aps.org/doi/10.1103/PhysRevB.49.12975}
}

@article{Haussmann2007,
  title = {Thermodynamics of the BCS-BEC crossover},
  author = {Haussmann, R. and Rantner, W. and Cerrito, S. and Zwerger, W.},
  journal = {Phys. Rev. A},
  volume = {75},
  issue = {2},
  pages = {023610},
  numpages = {22},
  year = {2007},
  month = {Feb},
  publisher = {American Physical Society},
  doi = {10.1103/PhysRevA.75.023610},
  url = {https://link.aps.org/doi/10.1103/PhysRevA.75.023610}
}

@article{Bloch2008,
  title = {Many-body physics with ultracold gases},
  author = {Bloch, Immanuel and Dalibard, Jean and Zwerger, Wilhelm},
  journal = {Rev. Mod. Phys.},
  volume = {80},
  issue = {3},
  pages = {885--964},
  numpages = {0},
  year = {2008},
  month = {Jul},
  publisher = {American Physical Society},
  doi = {10.1103/RevModPhys.80.885},
  url = {https://link.aps.org/doi/10.1103/RevModPhys.80.885}
}

@article{Vlietinck2013,
  title = {Quasiparticle properties of an impurity in a Fermi gas},
  author = {Vlietinck, Jonas and Ryckebusch, Jan and Van Houcke, Kris},
  journal = {Phys. Rev. B},
  volume = {87},
  issue = {11},
  pages = {115133},
  numpages = {11},
  year = {2013},
  month = {Mar},
  publisher = {American Physical Society},
  doi = {10.1103/PhysRevB.87.115133},
  url = {https://link.aps.org/doi/10.1103/PhysRevB.87.115133}
}

@article{Houcke2020,
  title = {High-precision numerical solution of the Fermi polaron problem and large-order behavior of its diagrammatic series},
  author = {Van Houcke, Kris and Werner, F\'elix and Rossi, Riccardo},
  journal = {Phys. Rev. B},
  volume = {101},
  issue = {4},
  pages = {045134},
  numpages = {14},
  year = {2020},
  month = {Jan},
  publisher = {American Physical Society},
  doi = {10.1103/PhysRevB.101.045134},
  url = {https://link.aps.org/doi/10.1103/PhysRevB.101.045134}
}

@article{Chien2010,
  title = {Comparative study of BCS-BEC crossover theories above ${T}_{c}$: The nature of the pseudogap in ultracold atomic Fermi gases},
  author = {Chien, Chih-Chun and Guo, Hao and He, Yan and Levin, K.},
  journal = {Phys. Rev. A},
  volume = {81},
  issue = {2},
  pages = {023622},
  numpages = {15},
  year = {2010},
  month = {Feb},
  publisher = {American Physical Society},
  doi = {10.1103/PhysRevA.81.023622},
  url = {https://link.aps.org/doi/10.1103/PhysRevA.81.023622}
}

@Article{Randeria2010,
author={Randeria, Mohit},
title={Pre-pairing for condensation},
journal={Nature Physics},
year={2010},
month={Aug},
day={01},
volume={6},
number={8},
pages={561-562},
issn={1745-2481},
doi={10.1038/nphys1748},
url={https://doi.org/10.1038/nphys1748}
}

@article{Dagotto1994,
  title = {Correlated electrons in high-temperature superconductors},
  author = {Dagotto, Elbio},
  journal = {Rev. Mod. Phys.},
  volume = {66},
  issue = {3},
  pages = {763--840},
  numpages = {0},
  year = {1994},
  month = {Jul},
  publisher = {American Physical Society},
  doi = {10.1103/RevModPhys.66.763},
  url = {https://link.aps.org/doi/10.1103/RevModPhys.66.763}
}

@article{Diener2008,
  title = {Quantum fluctuations in the superfluid state of the BCS-BEC crossover},
  author = {Diener, Roberto B. and Sensarma, Rajdeep and Randeria, Mohit},
  journal = {Phys. Rev. A},
  volume = {77},
  issue = {2},
  pages = {023626},
  numpages = {21},
  year = {2008},
  month = {Feb},
  publisher = {American Physical Society},
  doi = {10.1103/PhysRevA.77.023626},
  url = {https://link.aps.org/doi/10.1103/PhysRevA.77.023626}
}

@article{Milczewski2022,
  title = {Functional-renormalization-group approach to strongly coupled Bose-Fermi mixtures in two dimensions},
  author = {von Milczewski, Jonas and Rose, F\'elix and Schmidt, Richard},
  journal = {Phys. Rev. A},
  volume = {105},
  issue = {1},
  pages = {013317},
  numpages = {26},
  year = {2022},
  month = {Jan},
  publisher = {American Physical Society},
  doi = {10.1103/PhysRevA.105.013317},
  url = {https://link.aps.org/doi/10.1103/PhysRevA.105.013317}
}

@article{Pini2019,
  title = {Fermi gas throughout the BCS-BEC crossover: Comparative study of $t$-matrix approaches with various degrees of self-consistency},
  author = {Pini, M. and Pieri, P. and Strinati, G. Calvanese},
  journal = {Phys. Rev. B},
  volume = {99},
  issue = {9},
  pages = {094502},
  numpages = {19},
  year = {2019},
  month = {Mar},
  publisher = {American Physical Society},
  doi = {10.1103/PhysRevB.99.094502},
  url = {https://link.aps.org/doi/10.1103/PhysRevB.99.094502}
}

@Article{VanHoucke2012,
author={Van Houcke, K.
and Werner, F.
and Kozik, E.
and Prokof'ev, N.
and Svistunov, B.
and Ku, M. J. H.
and Sommer, A. T.
and Cheuk, L. W.
and Schirotzek, A.
and Zwierlein, M. W.},
title={Feynman diagrams versus Fermi-gas Feynman emulator},
journal={Nature Physics},
year={2012},
month={May},
day={01},
volume={8},
number={5},
pages={366-370},
issn={1745-2481},
doi={10.1038/nphys2273},
url={https://doi.org/10.1038/nphys2273}
}

@article{Rossi2018b,
  title = {Resummation of Diagrammatic Series with Zero Convergence Radius for Strongly Correlated Fermions},
  author = {Rossi, R. and Ohgoe, T. and Van Houcke, K. and Werner, F.},
  journal = {Phys. Rev. Lett.},
  volume = {121},
  issue = {13},
  pages = {130405},
  numpages = {6},
  year = {2018},
  month = {Sep},
  publisher = {American Physical Society},
  doi = {10.1103/PhysRevLett.121.130405},
  url = {https://link.aps.org/doi/10.1103/PhysRevLett.121.130405}
}

@article{Gorkov1961,
  Title                    = {{C}ontribution to the {T}heory of {S}uperfluidity in an {I}mperfect {F}ermi {G}as},
  Author                   = {Gor'kov, L.P. and Melik-Barkhudarov, T.K.},
  Journal                  = {Zh. Eskp. Theor. Fiz.},
  Year                     = {1961},
  Pages                    = {1452},
  Volume                   = {40}
}

@article{Petrov2004,
  title = {Weakly Bound Dimers of Fermionic Atoms},
  author = {Petrov, D. S. and Salomon, C. and Shlyapnikov, G. V.},
  journal = {Phys. Rev. Lett.},
  volume = {93},
  issue = {9},
  pages = {090404},
  numpages = {4},
  year = {2004},
  month = {Aug},
  publisher = {American Physical Society},
  doi = {10.1103/PhysRevLett.93.090404},
  url = {https://link.aps.org/doi/10.1103/PhysRevLett.93.090404}
}

@article{Haines1988,
author = {Haines, G. V. and Jones, Alan G.},
title = {Logarithmic Fourier transformation},
journal = {Geophysical Journal},
publisher = {Royal Astronomical Society},
volume = {92},
number = {1},
pages = {171-178},
year = {1988},
doi = {https://doi.org/10.1111/j.1365-246X.1988.tb01131.x},
url = {https://onlinelibrary.wiley.com/doi/abs/10.1111/j.1365-246X.1988.tb01131.x}
}

@ARTICLE{FFT1965,
  author = {Cooley, James and Tukey, John},
  title = {{A}n {A}lgorithm for the {M}achine {C}alculation of {C}omplex {F}ourier
	{S}eries},
  journal = {Mathematics of Computation},
  year = {1965},
  volume = {19},
  pages = {297-301},
  number = {90}
}

@article{Lang2019,
  title = {Fast logarithmic Fourier-Laplace transform of nonintegrable functions},
  author = {Lang, Johannes and Frank, Bernhard},
  journal = {Phys. Rev. E},
  volume = {100},
  issue = {5},
  pages = {053302},
  numpages = {13},
  year = {2019},
  month = {Nov},
  publisher = {American Physical Society},
  doi = {10.1103/PhysRevE.100.053302},
  url = {https://link.aps.org/doi/10.1103/PhysRevE.100.053302}
}

@article{Haussmann2009,
  title = {Spectral functions and rf response of ultracold fermionic atoms},
  author = {Haussmann, R. and Punk, M. and Zwerger, W.},
  journal = {Phys. Rev. A},
  volume = {80},
  issue = {6},
  pages = {063612},
  numpages = {18},
  year = {2009},
  month = {Dec},
  publisher = {American Physical Society},
  doi = {10.1103/PhysRevA.80.063612},
  url = {https://link.aps.org/doi/10.1103/PhysRevA.80.063612}
}

@article{Norman1998,
  title = {Phenomenology of the low-energy spectral function in high-${T}_{c}$ superconductors},
  author = {Norman, M. R. and Randeria, M. and Ding, H. and Campuzano, J. C.},
  journal = {Phys. Rev. B},
  volume = {57},
  issue = {18},
  pages = {R11093--R11096},
  numpages = {0},
  year = {1998},
  month = {May},
  publisher = {American Physical Society},
  doi = {10.1103/PhysRevB.57.R11093},
  url = {https://link.aps.org/doi/10.1103/PhysRevB.57.R11093}
}

@article{Magierski2009,
  title = {Finite-Temperature Pairing Gap of a Unitary Fermi Gas by Quantum Monte Carlo Calculations},
  author = {Magierski, Piotr and Wlaz\l{}owski, Gabriel and Bulgac, Aurel and Drut, Joaqu\'{\i}n E.},
  journal = {Phys. Rev. Lett.},
  volume = {103},
  issue = {21},
  pages = {210403},
  numpages = {4},
  year = {2009},
  month = {Nov},
  publisher = {American Physical Society},
  doi = {10.1103/PhysRevLett.103.210403},
  url = {https://link.aps.org/doi/10.1103/PhysRevLett.103.210403}
}

@article{Schmidt2011,
  title = {Excitation spectra and rf response near the polaron-to-molecule transition from the functional renormalization group},
  author = {Schmidt, Richard and Enss, Tilman},
  journal = {Phys. Rev. A},
  volume = {83},
  issue = {6},
  pages = {063620},
  numpages = {14},
  year = {2011},
  month = {Jun},
  publisher = {American Physical Society},
  doi = {10.1103/PhysRevA.83.063620},
  url = {https://link.aps.org/doi/10.1103/PhysRevA.83.063620}
}

@article{Luttinger1960,
  title = {Ground-State Energy of a Many-Fermion System. II},
  author = {Luttinger, J. M. and Ward, J. C.},
  journal = {Phys. Rev.},
  volume = {118},
  issue = {5},
  pages = {1417--1427},
  numpages = {0},
  year = {1960},
  month = {Jun},
  publisher = {American Physical Society},
  doi = {10.1103/PhysRev.118.1417},
  url = {https://link.aps.org/doi/10.1103/PhysRev.118.1417}
}

@article{Baym1962,
  title = {Self-Consistent Approximations in Many-Body Systems},
  author = {Baym, Gordon},
  journal = {Phys. Rev.},
  volume = {127},
  issue = {4},
  pages = {1391--1401},
  numpages = {0},
  year = {1962},
  month = {Aug},
  publisher = {American Physical Society},
  doi = {10.1103/PhysRev.127.1391},
  url = {https://link.aps.org/doi/10.1103/PhysRev.127.1391}
}

@article{Cornwall1974,
  title = {Effective action for composite operators},
  author = {Cornwall, John M. and Jackiw, R. and Tomboulis, E.},
  journal = {Phys. Rev. D},
  volume = {10},
  issue = {8},
  pages = {2428--2445},
  numpages = {0},
  year = {1974},
  month = {Oct},
  publisher = {American Physical Society},
  doi = {10.1103/PhysRevD.10.2428},
  url = {https://link.aps.org/doi/10.1103/PhysRevD.10.2428}
}

@article{gdd,
title = {Fast computation of divided differences and parallel hermite interpolation},
journal = {Journal of Complexity},
volume = {5},
number = {4},
pages = {417-437},
year = {1989},
issn = {0885-064X},
doi = {https://doi.org/10.1016/0885-064X(89)90018-6},
url = {https://www.sciencedirect.com/science/article/pii/0885064X89900186},
author = {Ömer Eǧecioǧlu and E Gallopoulos and Çetin K Koç}
}

@article{Bryan1990,
author={Bryan, R. K.},
title={Maximum entropy analysis of oversampled data problems},
journal={European Biophysics Journal},
year={1990},
month={Apr},
day={01},
volume={18},
number={3},
pages={165-174},
issn={1432-1017},
doi={10.1007/BF02427376},
url={https://doi.org/10.1007/BF02427376}
}

@article{Jarrell1996,
title = {Bayesian inference and the analytic continuation of imaginary-time quantum Monte Carlo data},
journal = {Physics Reports},
volume = {269},
number = {3},
pages = {133-195},
year = {1996},
issn = {0370-1573},
doi = {https://doi.org/10.1016/0370-1573(95)00074-7},
url = {https://www.sciencedirect.com/science/article/pii/0370157395000747},
author = {Mark Jarrell and J.E. Gubernatis}
}

@book{Sivia2006,
  title={Data Analysis: A Bayesian Tutorial},
  author={Sivia, D. and Skilling, J.},
  isbn={9780198568315},
  lccn={2006284782},
  series={Oxford science publications},
  url={https://books.google.de/books?id=lYMSDAAAQBAJ},
  year={2006},
  publisher={OUP Oxford}
}

@article{Goulko2017,
  title = {Numerical analytic continuation: Answers to well-posed questions},
  author = {Goulko, Olga and Mishchenko, Andrey S. and Pollet, Lode and Prokof'ev, Nikolay and Svistunov, Boris},
  journal = {Phys. Rev. B},
  volume = {95},
  issue = {1},
  pages = {014102},
  numpages = {11},
  year = {2017},
  month = {Jan},
  publisher = {American Physical Society},
  doi = {10.1103/PhysRevB.95.014102},
  url = {https://link.aps.org/doi/10.1103/PhysRevB.95.014102}
}

@article{Nogaki2023,
author = {Nogaki, Kosuke and Shinaoka, Hiroshi},
title = {Bosonic Nevanlinna Analytic Continuation},
journal = {Journal of the Physical Society of Japan},
volume = {92},
number = {3},
pages = {035001},
year = {2023},
doi = {10.7566/JPSJ.92.035001},
URL = {https://doi.org/10.7566/JPSJ.92.035001},
eprint = {https://doi.org/10.7566/JPSJ.92.035001}
}

@book{Atkinson1989, edition={2}, title={An introduction to numerical analysis}, publisher={Wiley}, author={Atkinson, E. Kendall}, year={1989}}

@book{Phillips2003, edition={1}, title={Interpolation and Approximation by Polynomials}, publisher={Wiley}, author={George M. Phillips}, year={2003}}

@Article{Sommer2011,
author={Sommer, Ariel
and Ku, Mark
and Roati, Giacomo
and Zwierlein, Martin W.},
title={Universal spin transport in a strongly interacting Fermi gas},
journal={Nature},
year={2011},
month={Apr},
day={01},
volume={472},
number={7342},
pages={201-204},
issn={1476-4687},
doi={10.1038/nature09989},
url={https://doi.org/10.1038/nature09989}
}

@article{Ku2012,
author = {Mark J. H. Ku  and Ariel T. Sommer  and Lawrence W. Cheuk  and Martin W. Zwierlein },
title = {Revealing the Superfluid Lambda Transition in the Universal Thermodynamics of a Unitary Fermi Gas},
journal = {Science},
volume = {335},
number = {6068},
pages = {563-567},
year = {2012},
doi = {10.1126/science.1214987},
URL = {https://www.science.org/doi/abs/10.1126/science.1214987},
eprint = {https://www.science.org/doi/pdf/10.1126/science.1214987}}

@Article{Navon2010,
  Title                    = {{T}he {G}round {S}tate of a {F}ermi {G}as with {T}unable {I}nteractions},
  Author                   = {Navon, N. and Nascimb\`{e}ne, S. and Chevy, F. and Salomon, C.},
  Journal                  = {Science},
  Year                     = {2010},
  Pages                    = {729},
  Volume                   = {328}
}

@article{Mukherjee2019,
  title = {Spectral Response and Contact of the Unitary Fermi Gas},
  author = {Mukherjee, Biswaroop and Patel, Parth B. and Yan, Zhenjie and Fletcher, Richard J. and Struck, Julian and Zwierlein, Martin W.},
  journal = {Phys. Rev. Lett.},
  volume = {122},
  issue = {20},
  pages = {203402},
  numpages = {6},
  year = {2019},
  month = {May},
  publisher = {American Physical Society},
  doi = {10.1103/PhysRevLett.122.203402},
  url = {https://link.aps.org/doi/10.1103/PhysRevLett.122.203402}
}

@article{Patel2020,
author = {Parth B. Patel  and Zhenjie Yan  and Biswaroop Mukherjee  and Richard J. Fletcher  and Julian Struck  and Martin W. Zwierlein },
title = {Universal sound diffusion in a strongly interacting Fermi gas},
journal = {Science},
volume = {370},
number = {6521},
pages = {1222-1226},
year = {2020},
doi = {10.1126/science.aaz5756},
URL = {https://www.science.org/doi/abs/10.1126/science.aaz5756},
eprint = {https://www.science.org/doi/pdf/10.1126/science.aaz5756}}

@article{Yan2019,
  title = {Boiling a Unitary Fermi Liquid},
  author = {Yan, Zhenjie and Patel, Parth B. and Mukherjee, Biswaroop and Fletcher, Richard J. and Struck, Julian and Zwierlein, Martin W.},
  journal = {Phys. Rev. Lett.},
  volume = {122},
  issue = {9},
  pages = {093401},
  numpages = {6},
  year = {2019},
  month = {Mar},
  publisher = {American Physical Society},
  doi = {10.1103/PhysRevLett.122.093401},
  url = {https://link.aps.org/doi/10.1103/PhysRevLett.122.093401}
}

@article{Schirotzek2008,
  title = {Determination of the Superfluid Gap in Atomic Fermi Gases by Quasiparticle Spectroscopy},
  author = {Schirotzek, Andr\'e and Shin, Yong-il and Schunck, Christian H. and Ketterle, Wolfgang},
  journal = {Phys. Rev. Lett.},
  volume = {101},
  issue = {14},
  pages = {140403},
  numpages = {4},
  year = {2008},
  month = {Oct},
  publisher = {American Physical Society},
  doi = {10.1103/PhysRevLett.101.140403},
  url = {https://link.aps.org/doi/10.1103/PhysRevLett.101.140403}
}

@article{Shin2008,
author={Shin, Yong-il
and Schunck, Christian H.
and Schirotzek, Andr{\'e}
and Ketterle, Wolfgang},
title={Phase diagram of a two-component Fermi gas with resonant interactions},
journal={Nature},
year={2008},
month={Feb},
day={01},
volume={451},
number={7179},
pages={689-693},
issn={1476-4687},
doi={10.1038/nature06473},
url={https://doi.org/10.1038/nature06473}
}

@article{Stewart2008,
author={Stewart, J. T.
and Gaebler, J. P.
and Jin, D. S.},
title={Using photoemission spectroscopy to probe a strongly interacting Fermi gas},
journal={Nature},
year={2008},
month={Aug},
day={01},
volume={454},
number={7205},
pages={744-747},
issn={1476-4687},
doi={10.1038/nature07172},
url={https://doi.org/10.1038/nature07172}
}

@Article{Gaebler2010,
author={Gaebler, J. P.
and Stewart, J. T.
and Drake, T. E.
and Jin, D. S.
and Perali, A.
and Pieri, P.
and Strinati, G. C.},
title={Observation of pseudogap behaviour in a strongly interacting Fermi gas},
journal={Nature Physics},
year={2010},
month={Aug},
day={01},
volume={6},
number={8},
pages={569-573},
issn={1745-2481},
doi={10.1038/nphys1709},
url={https://doi.org/10.1038/nphys1709}
}

@article{Feld2011,
author={Feld, Michael
and Fr{\"o}hlich, Bernd
and Vogt, Enrico
and Koschorreck, Marco
and K{\"o}hl, Michael},
title={Observation of a pairing pseudogap in a two-dimensional Fermi gas},
journal={Nature},
year={2011},
month={Dec},
day={01},
volume={480},
number={7375},
pages={75-78},
issn={1476-4687},
doi={10.1038/nature10627},
url={https://doi.org/10.1038/nature10627}
}

@misc{Li2023,
      title={Observation and quantification of pseudogap in unitary Fermi gases}, 
      author={Xi Li and Shuai Wang and Xiang Luo and Yu-Yang Zhou and Ke Xie and Hong-Chi Shen and Yu-Zhao Nie and Qijin Chen and Hui Hu and Yu-Ao Chen and Xing-Can Yao and Jian-Wei Pan},
      year={2023},
      eprint={2310.14024},
      archivePrefix={arXiv},
      primaryClass={cond-mat.quant-gas}
}

@Article{Pruefer2018,
author={Pr{\"u}fer, Maximilian
and Kunkel, Philipp
and Strobel, Helmut
and Lannig, Stefan
and Linnemann, Daniel
and Schmied, Christian-Marcel
and Berges, J{\"u}rgen
and Gasenzer, Thomas
and Oberthaler, Markus K.},
title={Observation of universal dynamics in a spinor Bose gas far from equilibrium},
journal={Nature},
year={2018},
month={Nov},
day={01},
volume={563},
number={7730},
pages={217-220},
issn={1476-4687},
doi={10.1038/s41586-018-0659-0},
url={https://doi.org/10.1038/s41586-018-0659-0}
}

@Article{Eigen2018,
author={Eigen, Christoph
and Glidden, Jake A. P.
and Lopes, Raphael
and Cornell, Eric A.
and Smith, Robert P.
and Hadzibabic, Zoran},
title={Universal prethermal dynamics of Bose gases quenched to unitarity},
journal={Nature},
year={2018},
month={Nov},
day={01},
volume={563},
number={7730},
pages={221-224},
issn={1476-4687},
doi={10.1038/s41586-018-0674-1},
url={https://doi.org/10.1038/s41586-018-0674-1}
}

@Article{Erne2018,
author={Erne, Sebastian
and B{\"u}cker, Robert
and Gasenzer, Thomas
and Berges, J{\"u}rgen
and Schmiedmayer, J{\"o}rg},
title={Universal dynamics in an isolated one-dimensional Bose gas far from equilibrium},
journal={Nature},
year={2018},
month={Nov},
day={01},
volume={563},
number={7730},
pages={225-229},
issn={1476-4687},
doi={10.1038/s41586-018-0667-0},
url={https://doi.org/10.1038/s41586-018-0667-0}
}

@article{scazza2017repulsive,
  title={Repulsive {F}ermi polarons in a resonant mixture of ultracold {Li}-6 atoms},
  author={Scazza, F and Valtolina, G and Massignan, P and Recati, Alessio and Amico, A and Burchianti, A and Fort, C and Inguscio, M and Zaccanti, M and Roati, G},
  journal={Phys Rev. Lett.},
  volume={118},
  number={8},
  pages={083602},
  year={2017},
  publisher={APS}
}

@article{Son2006,
  title = {Phase diagram of a cold polarized Fermi gas},
  author = {Son, D. T. and Stephanov, M. A.},
  journal = {Phys. Rev. A},
  volume = {74},
  issue = {1},
  pages = {013614},
  numpages = {6},
  year = {2006},
  month = {Jul},
  publisher = {American Physical Society},
  doi = {10.1103/PhysRevA.74.013614},
  url = {https://link.aps.org/doi/10.1103/PhysRevA.74.013614}
}

@article{Giorgini2008,
  title = {Phase Separation in a Polarized Fermi Gas at Zero Temperature},
  author = {Pilati, S. and Giorgini, S.},
  journal = {Phys. Rev. Lett.},
  volume = {100},
  issue = {3},
  pages = {030401},
  numpages = {4},
  year = {2008},
  month = {Jan},
  publisher = {American Physical Society},
  doi = {10.1103/PhysRevLett.100.030401},
  url = {https://link.aps.org/doi/10.1103/PhysRevLett.100.030401}
}

@article{Braaten2006,
title = {Universality in few-body systems with large scattering length},
journal = {Physics Reports},
volume = {428},
number = {5},
pages = {259-390},
year = {2006},
issn = {0370-1573},
doi = {https://doi.org/10.1016/j.physrep.2006.03.001},
url = {https://www.sciencedirect.com/science/article/pii/S0370157306000822},
author = {Eric Braaten and H.-W. Hammer}
}

@article{Braaten2010,
  title = {Short-Time Operator Product Expansion for rf Spectroscopy of a Strongly Interacting Fermi Gas},
  author = {Braaten, Eric and Kang, Daekyoung and Platter, Lucas},
  journal = {Phys. Rev. Lett.},
  volume = {104},
  issue = {22},
  pages = {223004},
  numpages = {4},
  year = {2010},
  month = {Jun},
  publisher = {American Physical Society},
  doi = {10.1103/PhysRevLett.104.223004},
  url = {https://link.aps.org/doi/10.1103/PhysRevLett.104.223004}
}

@article{Klein2020,
  title = {Normal State Properties of Quantum Critical Metals at Finite Temperature},
  author = {Klein, Avraham and Chubukov, Andrey V. and Schattner, Yoni and Berg, Erez},
  journal = {Phys. Rev. X},
  volume = {10},
  issue = {3},
  pages = {031053},
  numpages = {18},
  year = {2020},
  month = {Sep},
  publisher = {American Physical Society},
  doi = {10.1103/PhysRevX.10.031053},
  url = {https://link.aps.org/doi/10.1103/PhysRevX.10.031053}
}

@misc{FrankPiazza2020,
      title={$\omega/T$ scaling and IR/UV-mixing in Ising-nematic quantum critical metals}, 
      author={Bernhard Frank and Francesco Piazza},
      year={2020},
      eprint={2011.07076},
      archivePrefix={arXiv},
      primaryClass={cond-mat.str-el}
}

@article{LeeSS2017,
author = {Lee, Sung-Sik},
title = {Recent Developments in Non-Fermi Liquid Theory},
journal = {Annual Review of Condensed Matter Physics},
volume = {9},
number = {1},
pages = {227-244},
year = {2018},
doi = {10.1146/annurev-conmatphys-031016-025531},

URL = { 
    
        https://doi.org/10.1146/annurev-conmatphys-031016-025531
    
    

},
eprint = { 
    
        https://doi.org/10.1146/annurev-conmatphys-031016-025531
    
    

}
}

@article{Frank2023,
  title = {Marginal Fermi liquid at magnetic quantum criticality from dimensional confinement},
  author = {Frank, Bernhard and Liu, Zi Hong and Assaad, Fakher F. and Vojta, Matthias and Janssen, Lukas},
  journal = {Phys. Rev. B},
  volume = {108},
  issue = {10},
  pages = {L100405},
  numpages = {6},
  year = {2023},
  month = {Sep},
  publisher = {American Physical Society},
  doi = {10.1103/PhysRevB.108.L100405},
  url = {https://link.aps.org/doi/10.1103/PhysRevB.108.L100405}
}

@article{Loehneisen2007,
  title = {Fermi-liquid instabilities at magnetic quantum phase transitions},
  author = {L\"ohneysen, Hilbert v. and Rosch, Achim and Vojta, Matthias and W\"olfle, Peter},
  journal = {Rev. Mod. Phys.},
  volume = {79},
  issue = {3},
  pages = {1015--1075},
  numpages = {0},
  year = {2007},
  month = {Aug},
  publisher = {American Physical Society},
  doi = {10.1103/RevModPhys.79.1015},
  url = {https://link.aps.org/doi/10.1103/RevModPhys.79.1015}
}

@misc{diessel2022gMol,
      title={Probing molecular spectral functions and unconventional pairing using Raman spectroscopy}, 
      author={Oriana K. Diessel and Jonas von Milczewski and Arthur Christianen and Richard Schmidt},
      year={2022},
      eprint={2209.11758},
      archivePrefix={arXiv},
      primaryClass={cond-mat.quant-gas}
}

@article{Ryo2014a,
author = {Hanai, Ryo and Ohashi, Yoji},
year = {2014},
month = {04},
pages = {},
title = {Self-Consistent T-Matrix Approach to an Interacting Ultracold Fermi Gas with Mass Imbalance},
volume = {175},
journal = {Journal of Low Temperature Physics},
doi = {10.1007/s10909-013-0909-3}
}

@article{Ryo2014b,
  title = {Heteropairing and component-dependent pseudogap phenomena in an ultracold Fermi gas with different species with different masses},
  author = {Hanai, Ryo and Ohashi, Yoji},
  journal = {Phys. Rev. A},
  volume = {90},
  issue = {4},
  pages = {043622},
  numpages = {11},
  year = {2014},
  month = {Oct},
  publisher = {American Physical Society},
  doi = {10.1103/PhysRevA.90.043622},
  url = {https://link.aps.org/doi/10.1103/PhysRevA.90.043622}
}

@article{Pini2023,
  title = {Evolution of an attractive polarized Fermi gas: From a Fermi liquid of polarons to a non-Fermi liquid at the Fulde-Ferrell-Larkin-Ovchinnikov quantum critical point},
  author = {Pini, M. and Pieri, P. and Calvanese Strinati, G.},
  journal = {Phys. Rev. B},
  volume = {107},
  issue = {5},
  pages = {054505},
  numpages = {22},
  year = {2023},
  month = {Feb},
  publisher = {American Physical Society},
  doi = {10.1103/PhysRevB.107.054505},
  url = {https://link.aps.org/doi/10.1103/PhysRevB.107.054505}
}

@article{Pini2021,
  title = {Beyond-mean-field description of a trapped unitary Fermi gas with mass and population imbalance},
  author = {Pini, M. and Pieri, P. and Grimm, R. and Strinati, G. Calvanese},
  journal = {Phys. Rev. A},
  volume = {103},
  issue = {2},
  pages = {023314},
  numpages = {10},
  year = {2021},
  month = {Feb},
  publisher = {American Physical Society},
  doi = {10.1103/PhysRevA.103.023314},
  url = {https://link.aps.org/doi/10.1103/PhysRevA.103.023314}
}

@article{Perali2002,
  title = {Pseudogap and spectral function from superconducting fluctuations to the bosonic limit},
  author = {Perali, A. and Pieri, P. and Strinati, G. C. and Castellani, C.},
  journal = {Phys. Rev. B},
  volume = {66},
  issue = {2},
  pages = {024510},
  numpages = {16},
  year = {2002},
  month = {Jul},
  publisher = {American Physical Society},
  doi = {10.1103/PhysRevB.66.024510},
  url = {https://link.aps.org/doi/10.1103/PhysRevB.66.024510}
}

@article{Shunji2009,
  title = {Single-particle properties and pseudogap effects in the BCS-BEC crossover regime of an ultracold Fermi gas above ${T}_{c}$},
  author = {Tsuchiya, Shunji and Watanabe, Ryota and Ohashi, Yoji},
  journal = {Phys. Rev. A},
  volume = {80},
  issue = {3},
  pages = {033613},
  numpages = {9},
  year = {2009},
  month = {Sep},
  publisher = {American Physical Society},
  doi = {10.1103/PhysRevA.80.033613},
  url = {https://link.aps.org/doi/10.1103/PhysRevA.80.033613}
}

@misc{Pisani2023,
      title={Peaks and widths of radio-frequency spectra: An analysis of the phase diagram of ultra-cold Fermi gases}, 
      author={L. Pisani and M. Pini and P. Pieri and G. Calvanese Strinati},
      year={2023},
      eprint={2311.00479},
      archivePrefix={arXiv},
      primaryClass={cond-mat.quant-gas}
}

@misc{data,
title = "Data for figures",
  author = "C. H. Johansen and B. Frank and J. Lang",
  howpublished = "https://zenodo.org/records/10115898"
}
\end{document}